\documentclass[
  journal=pasa,
  manuscript=research-paper, 
  year=2026,
  volume=YY,
]{cup-journal}

\usepackage{amsmath}
\usepackage{amssymb}
\usepackage[table]{xcolor}
\usepackage{multirow}
\RequirePackage{aas-macros}
\usepackage{orcidlink}
\usepackage{savesym}
\savesymbol{vert}
\usepackage{yhmath}
\restoresymbol{SYM}{vert}
\usepackage{mathtools}
\usepackage[nopatch]{microtype}
\usepackage{booktabs}
\usepackage{listings}
\usepackage{graphicx}
\usepackage{graphbox} 
\usepackage{natbib}
\usepackage{hyperref}
\DeclareUnicodeCharacter{2212}{-}
\usepackage{lineno}
\usepackage{booktabs}
\usepackage{tabularx}
\usepackage{array}
\usepackage{rotating}
\usepackage{xcolor}

\defcitealias{VanKempen2024}{Paper~I}

\newcolumntype{C}{>{\centering\arraybackslash}X}

\title{The Galactic Squeeze: How Aggregate and Highly Dynamical Environments Shape Star Formation in the Local Universe}

\author{Wesley Van Kempen \orcidlink{0009-0009-8499-1326}}
\affiliation{Centre for Astrophysics and Supercomputing, Swinburne University of Technology, John Street, Hawthorn, VIC 3122, Australia}
\email[Wesley Van Kempen]{wvankempen@swin.edu.au}

\author{Michelle E. Cluver \orcidlink{0000-0002-9871-6490}}
\affiliation{Centre for Astrophysics and Supercomputing, Swinburne University of Technology, John Street, Hawthorn, VIC 3122, Australia}

\author{Edward N. Taylor \orcidlink{0000-0002-5522-9107}}
\affiliation{Centre for Astrophysics and Supercomputing, Swinburne University of Technology, John Street, Hawthorn, VIC 3122, Australia}

\author{Darren J. Croton \orcidlink{0000-0002-5009-512X}}
\affiliation{Centre for Astrophysics and Supercomputing, Swinburne University of Technology, John Street, Hawthorn, VIC 3122, Australia}
\alsoaffiliation{ARC Centre of Excellence for All Sky Astrophysics in 3 Dimensions (ASTRO 3D), Melbourne, Australia}
\alsoaffiliation{ARC Centre of Excellence for Dark Matter Particle Physics (CDM), Melbourne, Australia}

\author{Trystan S. Lambert \orcidlink{0000-0001-6263-0970}}
\affiliation{International Centre for Radio Astronomy Research (ICRAR), M468, University of Western Australia, 35 Stirling Hwy, Crawley, WA 6009, Australia}

\received {dd Mmm YYYY}
\revised  {dd Mmm YYYY}
\accepted {dd Mmm YYYY}
\published{XX XX 202X}


\begin{document}

\begin{abstract}
We investigate how galaxy evolution varies with environment in the nearby Universe by comparing an ``average'' reference volume in the Southern Galactic Pole (SGP) dataset from \citet{VanKempen2024} to the Nexus region, a dynamically assembling superstructure centred on the Abell~4038 galaxy cluster. We quantify environmental effects using the quenched fraction ($f_{\mathrm{Q}}$) and the mean and scatter of the specific star formation rate (sSFR) for the star-forming population, measured as functions of stellar mass and improved group-scale halo mass estimates from \cite{VanKempen2026}. One-dimensional binned trends are summarised with logistic fits for $f_{\mathrm{Q}}$ and a power-law describes the binned mean $\log{\rm sSFR}$ trends. We also de-couple the stellar--halo mass dependence in both the $f_{\mathrm{Q}}$ and mean $\log{\rm sSFR}$ analyses, demonstrating a joint dependence: $f_{\mathrm{Q}}$ increases with stellar mass in both field and group environments, while group galaxies show an additional dependence on halo mass. The Nexus exhibits systematic differences relative to the SGP baseline, consistent with increased heterogeneity in accretion histories and pre-processing within a forming superstructure. For star-forming galaxies, the mean $\log{\rm sSFR}$ declines strongly with stellar mass and shows additional environment-linked suppression in group-scale halos, while the scatter in $\log{\rm sSFR}$ varies primarily with stellar mass and shows comparatively weaker dependence on halo mass. However, these differences are generally within the uncertainties, and larger samples of dynamically evolving Nexus-like structures are required to determine whether they reflect genuine environmental effects or cosmic variance. Within the Nexus, splitting the sample into three projected radial zones around Abell~4038 shows that both quenching and the properties of the star-forming population vary systematically with distance from the node and forming supercluster, largely driven by differences in the sampled halo mass function, indicating that environmental regulation is not spatially uniform across the structure. Finally, a projected phase-space (PPS) analysis of Abell~4038 shows higher $f_{\mathrm{Q}}$ in regions associated with earlier infall, linking quenching to orbital history within the cluster. However, when grouping infall PPS zones and splitting by stellar mass, this trend is strongly mass dependent, with low-mass galaxies ($\log{M_{\rm{stellar}}}<10$) showing no significant change in $f_{\mathrm{Q}}$. These results demonstrate that the drivers of galaxy evolution depend jointly on stellar mass, the halo mass of the local group environment, and location within the surrounding large-scale structure. This motivates future, larger-statistics, multi-wavelength studies that combine wide-area spectroscopy with tracers of gas, dust, and hot halos to test quenching and star formation regulation mechanisms across the cosmic web.
\end{abstract}

\section{Introduction}
\label{sec:Introduction}

Galaxies are not isolated systems; rather, their properties, such as star formation rate (SFR), morphology, stellar mass, gas content, colour, kinematics, and structural parameters are strongly influenced by their environments \citep{Dressler1980, Kauffmann2004, Peng2010, Blanton2009}. The environment of galaxies encompasses a broad range of densities, both on small ($<1$ Mpc) and large (a few to tens of Mpc) scales. On small scales, galaxies span from low density environments, where close companions are rare, to those embedded in groups and the densely packed cores of galaxy clusters \citep{Eke2004, Yang2007, Robotham2011}. On large-scales, galaxies trace the cosmic web, a network of filaments, sheets, and voids that define the structure of the local Universe. The cosmic web arises naturally within the $\Lambda$CDM paradigm \citep{Davis1985, Springel2005, Planck2016}, where gravitational growth from primordial density perturbations leads to the accretion of dark matter and baryons, as described by the Zel’dovich approximation \citep{Zeldovich1970}. Understanding how environment shapes the baryon cycle is critical to studying the physical mechanisms that drive star formation and quenching accross cosmic epochs \citep{Peng2010, Davies2019, Cluver2020, Bluck2020}.

Galaxy clusters represent some of the most extreme environments in the Universe and show significant differences in the properties of their member galaxies. The high-density intracluster medium (ICM) strips cold gas reservoirs through ram pressure stripping, which suppresses star formation on short timescales \citep{Gunn1972, Abadi1999, Jaffe2015, Boselli2022}. Tidal interactions with the cluster potential and frequent galaxy–galaxy encounters and mergers accelerate morphological transformation and promote the build-up of spheroidal systems \citep{Moore1996, Boselli2006, Joshi2020}. The cluster environment also incorporates pre-processing, where galaxies experience environmental quenching within infalling groups before reaching the cluster core \citep{Zabludoff1998, Fujita2004, Haines2015, Hou2014}. These processes collectively drive the well-established morphology–density and star formation–density relations, where passive early-type galaxies dominate cluster cores, while star-forming disk galaxies are preferentially found in lower-density outskirts \citep{Dressler1980, Bamford2009, Peng2010, Wetzel2012}. Understanding the relative contributions of these mechanisms is essential for disentangling the role of clusters in shaping galaxy evolution relative to other large-scale environmental influences.

Forming superclusters, in contrast to virialised clusters, are highly dynamical and unrelaxed environments. A significant portion of their mass is still assembling through the merging of clusters, groups, and infalling field galaxies, which creates a complex and evolving gravitational landscape \citep{Einasto2014, Chon2015, Owers2013}. Superclusters are the largest gravitationally bound structures in the Universe and mark the final stages of large-scale structure formation. They provide a rich framework for studying how environmental processes shape galaxy evolution across cosmic time \citep[e.g.][]{Einasto2011, Luparello2013}. Studies of nearby forming superclusters, such as the Shapley Supercluster \citep[e.g.][]{Pimbblet2002, Haines2011, Owers2013} and the Saraswati Supercluster \citep[e.g.][]{Bagchi2017}, have shown evidence of enhanced star formation in galaxies along filaments and in infalling groups. This suggests that large-scale accretion may temporarily trigger star formation before quenching mechanisms take effect \citep{Porter2008, Mahajan2013, Luber2022}. As galaxies move deeper into the supercluster potential, environmental effects such as gas starvation, harassment, and merger-driven quenching become more prominent \citep{Porter2008, Mahajan2013, Luber2022}. Observations indicate that pre-processing in filaments and groups plays a major role in shaping galaxy properties before cluster infall, with quenching fractions increasing toward the supercluster core \citep{Cybulski2014, Sohn2021, Haines2015}.

In \citet{VanKempen2024} (hereafter \citetalias{VanKempen2024}), the role of small-scale environments in shaping galaxy evolution was investigated, with a focus on how the local environment of galaxies, specifically galaxy groups and the field modulate star formation. Using a mass-complete sample to ($z < 0.1$), it was found that galaxy groups exhibit a higher fraction of quenched galaxies, star-forming galaxies showed larger star formation deficits compared to the isolated field when controlling for stellar mass, and the quenching efficiency increased as a function group multiplicity. These results highlighted the importance of hierarchical environmental effects in regulating star formation, where galaxies experience pre-processing in group environments before entering larger structures such as clusters \citep{Wetzel2012, Hou2014, Davies2019}.

The dataset used in \citetalias{VanKempen2024} was drawn from the Southern Galactic Pole (SGP) region, a 376 deg\(^2\) field primarily composed of galaxies from the 2dF Galaxy Redshift Survey (2dFGRS; \citealt{Colless2001}) and the Galaxy And Mass Assembly (GAMA G23; \citealt{Driver2011, Driver2022}) surveys. This dataset provides a statistically robust sample of galaxies with well-characterised stellar masses and star formation rates, which enables detailed environmental studies. Group catalogues were constructed using a Friends-of-Friends (FoF) algorithm \citep{Eke2004, Robotham2011} and spectroscopically confirmed galaxy associations. This approach allowed \citetalias{VanKempen2024} to examine the impact of small-scale environments on galaxy properties while maintaining a well-defined field comparison sample.

Since the publication of \citetalias{VanKempen2024}, significant methodological advancements have been achieved in the estimation of group halo masses, as detailed in \citet{VanKempen2026}. In \citetalias{VanKempen2024}, group multiplicity was used as a proxy for environment, due to the substantial uncertainties in halo mass estimates (also detailed in \citet{VanKempen2026}), particularly for low-mass group-halos. In \citet{VanKempen2026}, two observational methods for halo mass estimation were developed and validated: a modified virial theorem approach and a summed stellar-to-halo mass relation derived using the S\textsc{hark} semi-analytic model \citep{Lagos2018, Lagos2024}, with cross-validation against both the \textsc{Semi-Analytic Galaxy Evolution} \citep[SAGE;][]{Croton2016} and the \textsc{Galaxy Evolution and Assembly} semi-analytic model \citep[GAEA;][]{DeLucia2014, Hirschmann2016, DeLucia2024}. For the purposes of this study, we use the summed stellar-to-halo mass relation (Equation 11 in \citet{VanKempen2026}) to estimate the halo masses for galaxy groups. The summed stellar-to-halo mass relation yields the most accurate group halo masses, especially in the low-mass regime (see \citealt{VanKempen2026} for full details on the halo mass estimators and their use cases). This methodological improvement enables a more robust characterisation of environmental effects across the full range of halo masses.

Building on these foundations, the present study expands the scope to a direct cross-comparison of environmental signatures on galaxy evolution, using halo mass as the principal proxy for environment. Specifically, the global distribution of star formation and quenching in the SGP dataset is contrasted with that observed in a highly dynamical forming supercluster region, ``the Nexus'', centred on Abell 4038 (hereafter A4038). This approach enables a systematic assessment of how the processes governing star formation and quenching differ between the global field and group population, and those operating within the unique gravitational landscape of a forming supercluster. The analysis incorporates a projected phase–space study of A4038, a comparative investigation of global quenching and star-forming trends between the SGP and Nexus regions, and a detailed examination of radial environmental effects within the Nexus, subdivided into three zones. These comparisons are facilitated by the improved halo mass estimates, allowing for a consistent and physically motivated environmental metric across all analyses.

This paper is structured as follows. Section~\ref{sec:Data and Sample Selection} describes the SGP dataset, the Nexus region, and the selection of the radial zones within the Nexus. Section~\ref{sec:Results} presents our results, including: (i) global quenching trends, (ii) global star formation trends, (iii) the radial impact on quenching and star formation within the Nexus, and (iv) a projected phase–space analysis of A4038. Section~\ref{sec:Discussion} discusses the implications of these findings and places them within the wider context of environmentally driven galaxy evolution. Section~\ref{sec:Sum_Con} summarises the main results and outlines prospects for future work.

Throughout this study, the cosmological parameters \( H_{0} = 70 \) km s\(^{-1}\) Mpc\(^{-1}\), \( \Omega_{M} = 0.3 \), and \( \Omega_{\Lambda} = 0.7 \) are adopted. A \citet{Chabrier2003} initial mass function (IMF) is assumed, and all dark matter halo masses, M$_\mathrm{halo}~(M_{\odot})$, are defined as $M_{200}$, the mass enclosed within a radius where the mean density is 200 times the critical density of the Universe, unless stated otherwise.

\section{Data and Sample Selection}
\label{sec:Data and Sample Selection}

\subsection{SGP Sample}
\label{subsec:SGP}

\begin{figure*}[t]
\centering
\includegraphics[width=\linewidth]{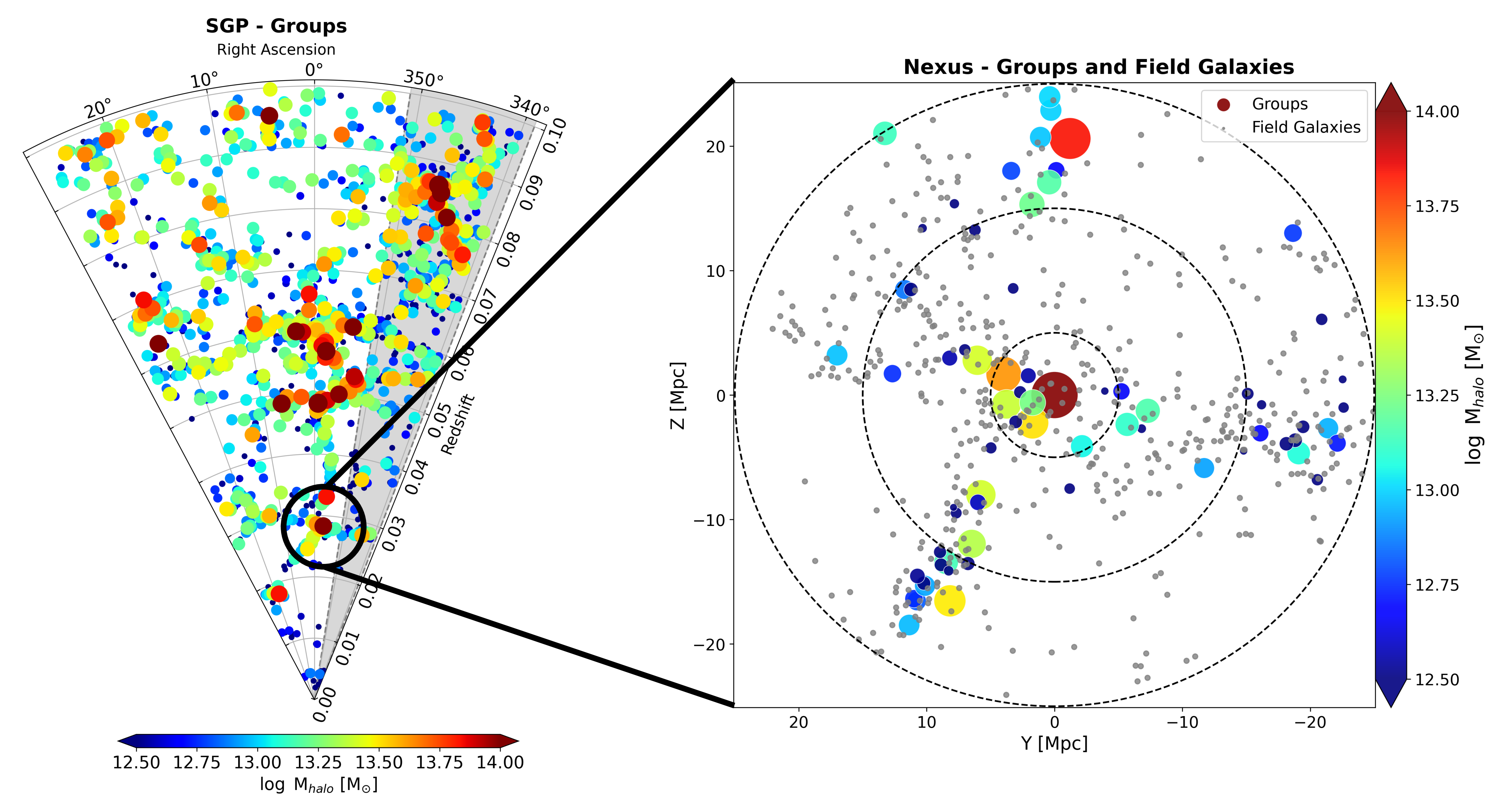}
\caption{The left panel indicates the distribution of galaxy groups in the larger SGP dataset from \citetalias{VanKempen2024}, highlighting the location of the extracted forming supercluster, the ``Nexus''. The grey shaded region indicates the Right Ascension boundaries of the GAMA G23 survey. The right panel presents the spatial distribution of galaxy groups in the Nexus region transformed into comoving Cartesian coordinates. The coordinate system has been rotated such that the line-of-sight distribution of the Abell~4038 cluster galaxies lies along the $Z$-axis. The $Y$-axis primarily traces Declination with a small contribution from Right Ascension due to this rotation. The dashed circles indicate projected radial zones of $0$--$5$~Mpc, $5$--$15$~Mpc, and $15$--$25$~Mpc from the centre of Abell~4038.}
\label{Nexus}
\end{figure*}

This work uses the SGP observational dataset, introduced in \citetalias{VanKempen2024}, which provides a comprehensive spectroscopic and photometric dataset of galaxies over a contiguous 376~deg$^2$ ($340^\circ < \mathrm{RA} < 26^\circ$, $-35.3^\circ < \mathrm{Dec} < -25.8^\circ$) volume to a limit of $z < 0.1$. This field overlaps and combines the 2dFGRS and GAMA G23 fields \citep{Colless2001, Driver2009}. The volume combines redshifts primarily from 2dFGRS and GAMA, but is also supplemented by 2-degree Field Lensing Survey \citep[2dFLenS;][]{Blake2016}, the 6-degree Field Galaxy Redshift Survey \citep[6dFGRS;][]{Jones2004}, the 2MASS Redshift Survey \citep[2MRS;][]{Macri2019}, and the Million Quasars catalogue \citep[MILLIQUAS;][]{Flesch2021}. This contiguous, heterogeneous sample consists of 24,656 unique spectroscopic sources, and $\sim$93\% of all sources (22,933 galaxies) have been cross-matched to photometry from the Wide-Field Infrared Survey Explorer \citep[WISE;][]{Wright2010}, providing the dataset with accurate measurements of both stellar mass \citep[$M_\mathrm{stellar}$;][]{Jarrett2023} and star formation rates \citep[SFR;][]{Cluver2025}.

Within the SGP dataset, galaxy groups were identified via a graph-theory derived Friends-of-Friends (FoF) algorithm \citep[\mbox{pyFoF}; ][]{Lambert2020}\footnote{pyFoF can be acsessed via: \url{https://github.com/TrystanScottLambert/pyFoF}}, see \citetalias{VanKempen2024} for full details on the optimisation of the FoF based finder. The final group catalogue included 1,413 unique groups containing 8,980 galaxies, of which 8,454 were successfully cross-matched with WISE. All halo mass estimates are derived from the summed stellar-to-halo mass relation (Equation 11 in \cite{VanKempen2026}), which details the relation between the summed stellar masses of the 3 most massive galaxies in a group to estimate the dark matter halo mass (M$_\mathrm{halo}$). This relation gives typical uncertainties of $\sim$0.15 dex in M$_\mathrm{halo}$.

\subsection{SAGE}
\label{subsec:SAGE}

The SAGE model \citep{Croton2016} is a publicly available semi-analytic framework that builds upon the Munich family of semi-analytic galaxy formation models originally developed for the Millennium Simulation \citep{Croton2006}, incorporating comprehensive treatments of radiative cooling, star formation, feedback from supernovae and AGN, black hole growth, and environmental processes. Model parameters are calibrated to reproduce key observables, including the stellar mass function and galaxy colour distributions at $z = 0$.

For this analysis, SAGE was implemented on the BOLSHOI $N$-body simulation \citep{Klypin2011}, which simulates a cosmological volume of $(250~h^{-1}~\mathrm{Mpc})^3$ with $2048^3$ particles, yielding a mass resolution of $1.35 \times 10^8~h^{-1}~M_{\odot}$. Halo identification used the \textsc{ROCKSTAR} phase-space halo finder \citep{Behroozi2013a}, and merger trees were constructed using the \textsc{Consistent Trees} algorithm \citep{Behroozi2013b}. The simulation adopts WMAP7 cosmological parameters \citep{Komatsu2011}: $\Omega_{\mathrm{m}} = 0.270$, $\Omega_{\Lambda} = 0.730$, $\Omega_{\mathrm{b}} = 0.0469$, $h = 0.70$, $\sigma_8 = 0.82$, and $n_{\mathrm{s}} = 0.95$.

In this work, a set of 30 SAGE mock light-cones was generated, each spanning a large area of $2,685~\mathrm{deg}^2$. To ensure consistency with the observational selection in the Nexus region, a Johnson $B$ band magnitude limit of $19.45$ (Vega) was imposed, matching the 2dFGRS survey completeness limit. These mock catalogues were primarily used to identify Nexus-like forming supercluster analogues and to evaluate the reliability of galaxy assignments to radial zones centred on the cluster centre. The selection of Nexus analogues was based on criteria including central cluster membership, redshift range, isolation from other massive structures, and the presence of neighbouring halos within specified mass intervals (see Section~\ref{subsec:nexus_zones} for details).

\subsection{The Nexus: A Forming Supercluster}
\label{subsec:nexus}

Superclusters trace the densest nodes of the cosmic web and are natural sites of accelerated assembly and complex dynamics \citep[e.g.][]{Einasto2011, Chon2013}. Star-formation activity is known to vary across supercluster environments and filaments—often enhanced in infall regions and suppressed toward the densest cores—providing a sensitive probe of environmental regulation in the nearby Universe \citep[e.g.][]{Porter2007, Haines2015}.

We define the Nexus as a spherical region of radius 25~Mpc centred on A4038. This volume contains 1216 galaxies, of which 696 are associated with groups (including A4038); 524 of these group members reside in groups outside A4038, leaving 520 field galaxies (i.e., galaxies not associated to any group). Within the Nexus there are eight groups with $M_{\rm halo}>10^{13}\,{\rm M_{\odot}}$, four of which lie within 5~Mpc of A4038, underscoring that this is a dynamically evolving cosmic-structure likely to assemble into a supercluster in the near future.

The left-hand panel of Figure~\ref{Nexus} shows the galaxy groups in the full SGP catalogue from \citetalias{VanKempen2024}, and highlights the location of the Nexus within the SGP region. The right-hand panel showcases the distribution of groups and field galaxies within the Nexus region; in this panel, we also introduce three radial zones ($0$–$5$~Mpc, $5$–$15$~Mpc, and $15$–$25$~Mpc from A4038) in which we will assess star formation trends as a function of cluster-centric distance to test how assembling superclusters affect star formation in the local Universe.

\subsection{Nexus Radial Zones}
\label{subsec:nexus_zones}

\begin{figure*}[!ht]
\centering
\includegraphics[width=0.49\textwidth]{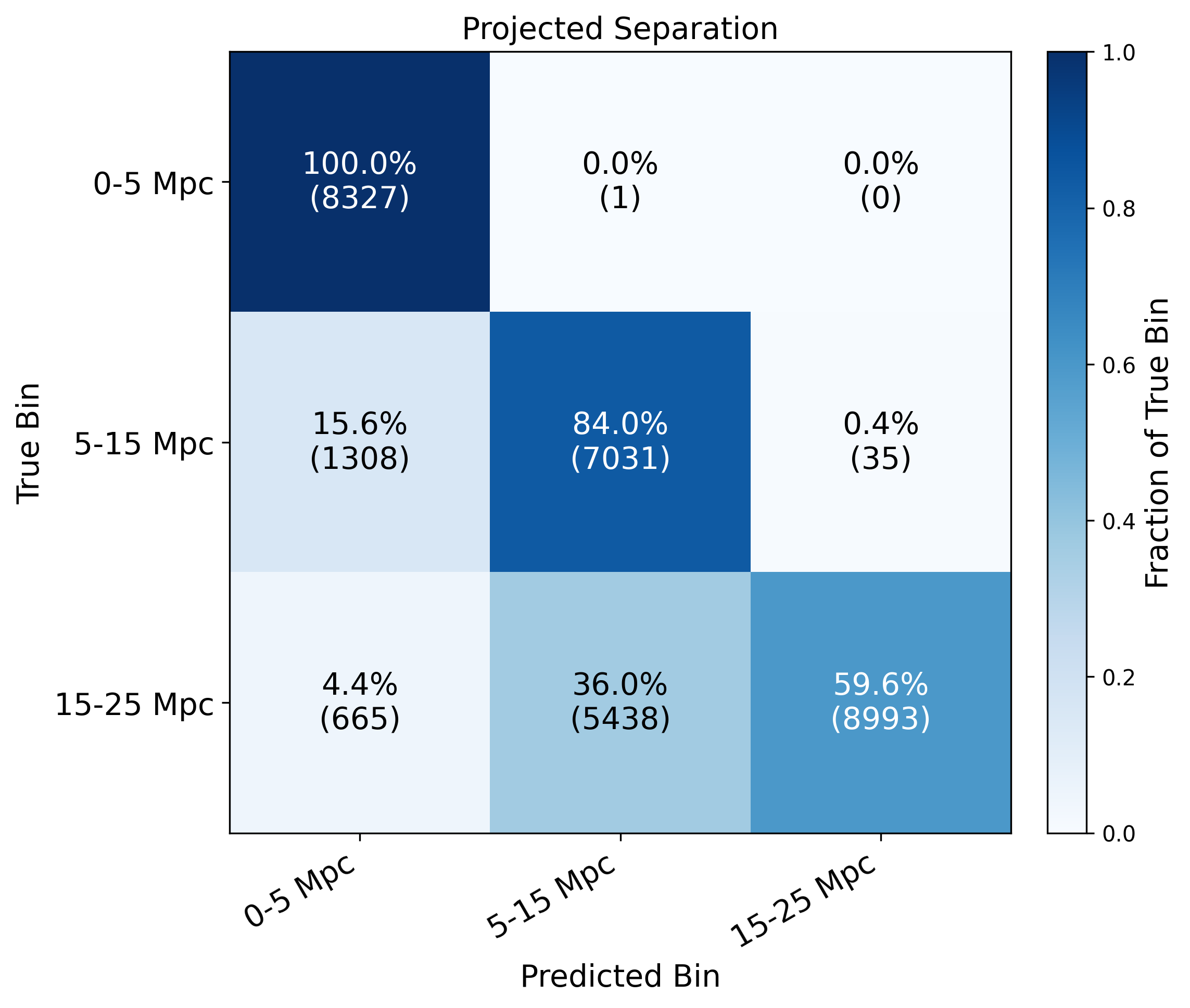}
\includegraphics[width=0.49\textwidth]{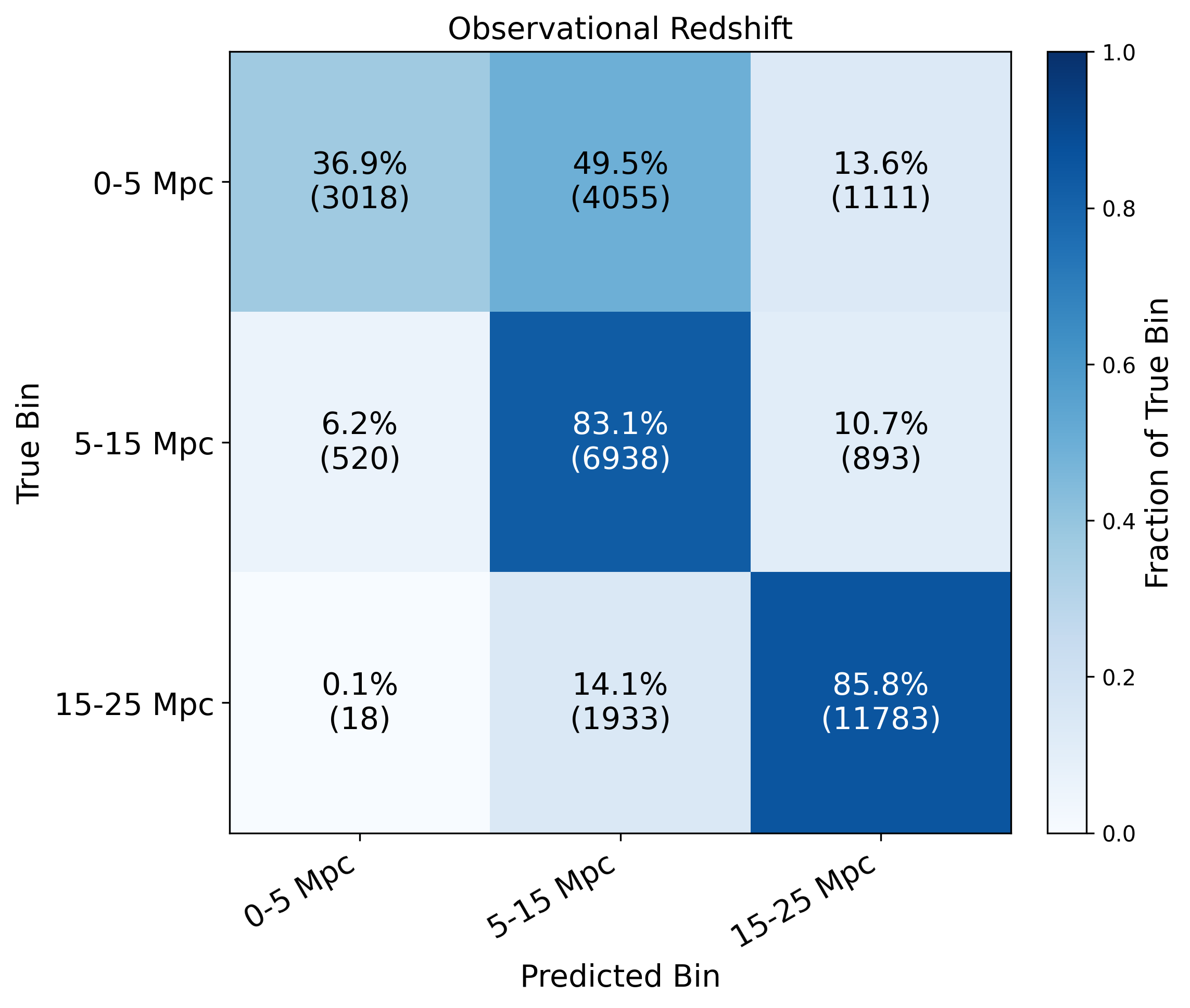}
\includegraphics[width=0.49\textwidth]{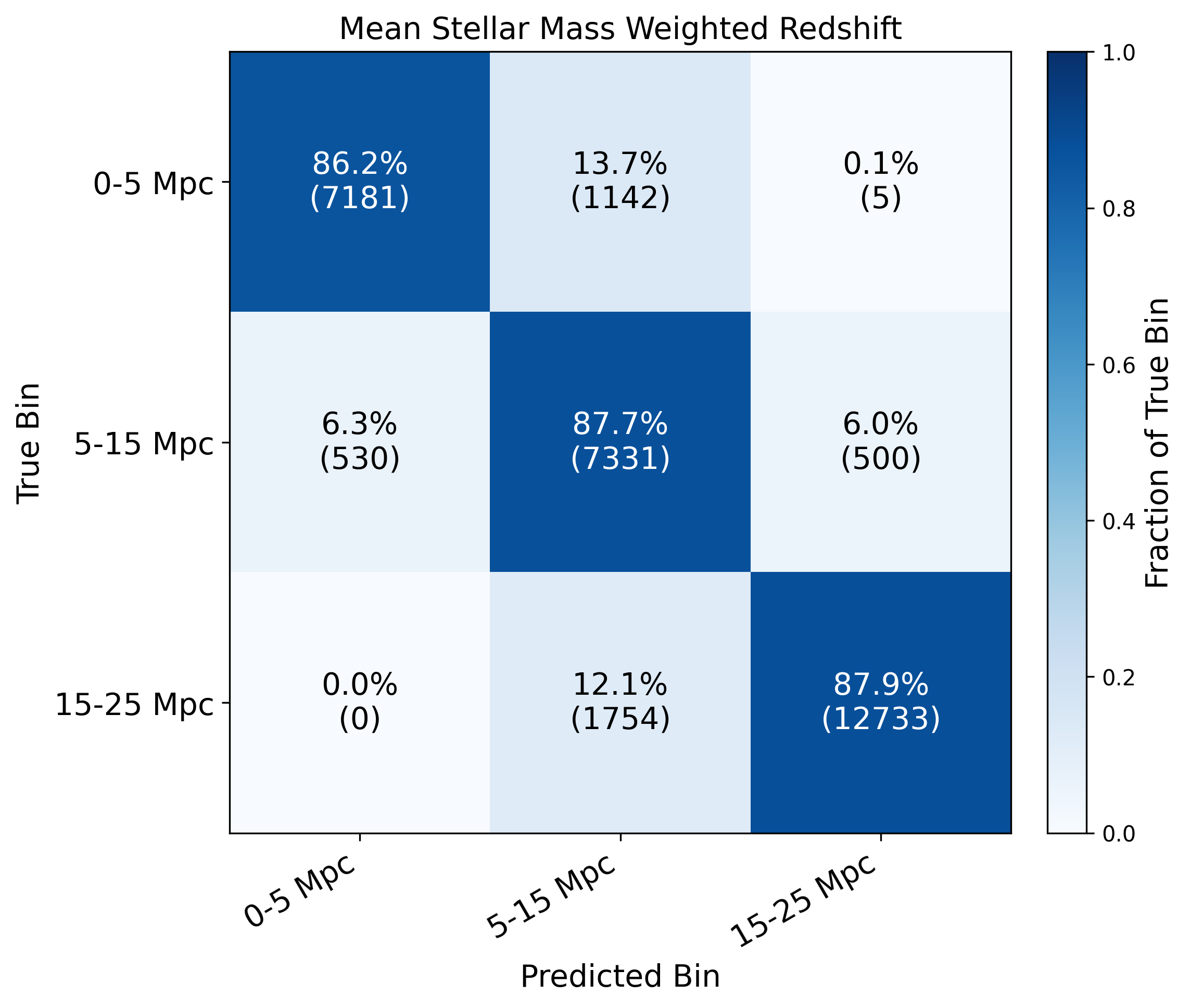}
\caption{Confusion matrices comparing predicted versus true distance bins from the centre of the large-scale structure, evaluated across all 40 Nexus-like analogues identified in the SAGE-Bolshoi lightcones. Each row corresponds to the true bin and each column to the predicted bin, with cell values indicating the fraction of galaxies in each true bin assigned to each predicted bin, with the raw galaxy counts given in parentheses. From left to right: projected separation, observational redshift (i.e. uncorrected three-dimensional distances), and stellar-mass–weighted mean redshift. The diagonal entries represent correctly classified galaxies.}
\label{fig:Zone_Uncertainty}
\end{figure*}

To assess environmental trends as a function of cluster-centric distance from A4038, three, three-dimensional comoving radial shells are defined: $0$–$5$~Mpc, $5$–$15$~Mpc, and $15$–$25$~Mpc, centred on A4038. Accurately inferring three-dimensional galaxy positions is inherently complicated by redshift-space distortions, which bias line-of-sight distances inferred from observed redshifts. On large scales, coherent infall towards overdensities produces an apparent compression of structures along the line-of-sight (the Kaiser effect; \citealt{Kaiser1987}), while on small scales, virial motions within haloes elongate galaxy distributions into ``Fingers of God'' \citep[FoG: ][see also \citealt{Hamilton1998}]{Jackson1972}. These effects can introduce errors of several Mpc in inferred distances and substantially complicate attempts to measure star-formation trends as a function of true three-dimensional position \citep[e.g.][]{Reid2009, Mamon2013, Oman2013}.

A common observational approach is to use projected separations from the cluster centre \citep[e.g.][]{Carlberg1997, Rines2003, Haines2015}. While this method avoids biases from redshift-space distortions, it neglects the line-of-sight dimension and is therefore susceptible to projection effects. This behaviour is quantified in Figure~\ref{fig:Zone_Uncertainty}, which presents a combined confusion matrix of 40 Nexus-like structures drawn from our SAGE light-cones, providing an estimate of the expected sample variance. Nexus analogues are selected to match the following criteria: the central cluster lies within $0.015 < z < 0.045$, has $130 < N_{\mathrm{members}} < 250$, no other comparable structure within 28~Mpc, and at least six neighbouring halos within 25~Mpc with $13.35 < \log(M_{\mathrm{halo}}) < 13.85$. For the projected separation method, the $5$–$15$ and $15$–$25$~Mpc bins seem to recover the true membership quite well ($100\%$ and $84\%$ respectively), they are subject to contamination from off-diagonal bins, and thus contain moderate leakage from non-adjacent bins e.g. $\sim19\%$ (1937 galaxies from the 10227 reported) for the $0$–$5$~Mpc bin and $\sim44\%$ (5439 galaxies from the 12470 reported) for the $5$–$15$~Mpc. For the $0$–$5$~Mpc bin, this corresponds to $\sim80.8\%$ of galaxies in the projected $0$–$5$~Mpc zone being true members and $\sim56.3\%$ for the $5$–$15$~Mpc zone. This contamination propagates into the outer bins, where both completeness and purity degrade. For example, in the $15$–$25$~Mpc bin, only $59.6\%$ of galaxies are correctly classified, with a substantial fraction originating from the $5$–$15$~Mpc bin. These results demonstrate that although projected separation yields high completeness in the innermost regions, it introduces non-negligible contamination that can bias environmental trends, particularly when interpreting gradients in galaxy properties with cluster-centric distance.

Alternatively, one may attempt to reconstruct full three-dimensional distances using observational redshifts. However, as shown in Figure~\ref{fig:Zone_Uncertainty}, this approach is strongly biased by redshift-space distortions. For example, only $36.9\%$ of galaxies in the true $0$–$5$~Mpc bin are recovered in the correct bin, with nearly half ($49.5\%$) scattered into the $5$–$15$~Mpc bin due to line-of-sight velocity distortions. Similar mixing is observed across all bins, indicating that naïvely using observational redshifts leads to significant radial smearing and systematic misclassification.

To mitigate these biases, we adopt a stellar-mass–weighted mean redshift to estimate the group centre along the line-of-sight. This approach effectively collapses each group to a single line-of-sight distance (inferred from the redshift), reducing the impact of internal velocity dispersion (FoG) while retaining sensitivity to large-scale structure. The weighted mean redshift for a group is computed as:
\begin{equation}
\overline{z}_{\mathrm{weighted}} = \frac{\sum_{i} w_{i} z_{i}}{\sum_{i} w_{i}},
\end{equation}
where $w_{i} = \mathrm{M}_{\mathrm{stellar},i} \,[M_{\odot}]$. This weighting emphasises the most massive group members, which are more likely to reside near the potential minimum and provide a more stable estimate of the group centre. Importantly, the true line-of-sight centre ($c\Delta z$) cannot be directly recovered from observational data due to redshift-space distortions; however, tests on the SAGE lightcones demonstrate that this estimator provides the closest statistical approximation to the true centre when compared to alternative metrics (e.g. unweighted mean or median redshift).

The effectiveness of this method is demonstrated in Figure~\ref{fig:Zone_Uncertainty}. The stellar-mass–weighted redshift recovers $86.2\%$, $87.7\%$, and $87.9\%$ of the true galaxies in the correct bins for the $0$–$5$, $5$–$15$, and $15$–$25$~Mpc zones, respectively. In contrast to both projected separation and observational redshift, the off-diagonal contamination is substantially reduced, with minimal leakage between non-adjacent bins e.g. $<7\%$ (530 galaxies from the 7711 reported) from $0$–$5$, $<29\%$ (2896 galaxies from the 10227 reported) from $5$–$15$~Mpc and $<4\%$ (505 galaxies from the reported 13,238) from $5$–$15$~Mpc. For the $0$–$5$~Mpc bin, this corresponds to $\sim93.1\%$ of galaxies in the projected $0$–$5$~Mpc zone being true members, $\sim71.7\%$ for the $5$–$15$~Mpc zone and $\sim96.2\%$ for the $15$–$25$~Mpc zone. This demonstrates that the weighted redshift method provides a significantly improved balance between completeness and purity across all radial zones.

These results highlight that the choice of distance metric can strongly bias environmental trends if not carefully considered. Projected separations systematically inflate inner regions with foreground and background interlopers, while observational redshifts introduce significant radial smearing due to peculiar velocities. The stellar-mass–weighted redshift approach mitigates both effects by anchoring the line-of-sight position to a physically motivated group centre, thereby providing the most reliable classification of galaxies into radial zones for this analysis.

Nevertheless, it is important to emphasise that no observational method can fully recover the true three-dimensional positions of galaxies in the presence of redshift-space distortions. Residual uncertainties remain due to large-scale flows and groups straddling bin boundaries. Our approach therefore aims to minimise, rather than eliminate, these biases. We advocate that similar correction schemes be adopted and further refined in future studies, particularly when investigating environmental trends as a function of cluster-centric distance, filamentary distance, or any other distance metric where systematic misclassification can significantly impact inferred physical correlations.

\subsection{Abell 4038}
\label{subsec:abell4038}

A4038 (RA = 357°03' 31'', Dec = −28°07' 38'', z = 0.029) is a rich southern galaxy cluster exhibiting clear signatures of recent dynamical activity \citep{Owers2013}. From the SGP group catalogue, A4038 has 172 member galaxies. The cluster hosts a compact radio relic with a projected extent of $\sim 56$~kpc and an exceptionally steep integrated spectral index, $\alpha \simeq -2.2$ \citep{Slee1998}. Steep-spectrum, diffuse synchrotron emission of this kind is widely interpreted as tracing relativistic electrons accelerated and/or re-energised by merger-driven shocks in the ICM, followed by spectral ageing due to synchrotron and inverse-Compton losses \citep[e.g.][]{Ensslin2001, Feretti2012, Brunetti2014, vanWeeren2019}. The spectrum in A4038 is therefore indicative of an advanced post-shock ageing stage, consistent with a merger episode within the past $\lesssim 1$ Gyr \citep{Ensslin2001, vanWeeren2019}.

Independent constraints place A4038 at the low-mass end of the local cluster population. Using the stellar–halo mass relation from \citet{VanKempen2026}, it estimates the halo mass to be $M_{200}=10^{14.0}\,{\rm M_{\odot}}$. This is comparable to the Sunyaev–Zel’dovich (SZ) estimated mass from \citet{Planck2016}, which yields $M_{500} = 10^{14.2}\,{\rm M_{\odot}}$. Adopting a standard NFW halo profile \citep{Navarro1997} and a plausible concentration–mass relation at low redshift to convert between spherical overdensities \citep{Hu2003}, the SZ estimate corresponds to $M_{200}\approx 10^{14.33}\,{\rm M_{\odot}}$. Given the uncertainties in both halo mass estimates and the uncertainty in the conversion between spherical over-densities, the two mass estimates are within agreement of each other. Taken together with the radio evidence for recent shock activity, A4038 is best described as a rich low-mass cluster in a dynamically evolving state. These properties make A4038 a compelling testbed for quantifying how dynamically evolving environments affect galaxy evolution.

\section{Results}
\label{sec:Results}


\subsection{Global Quenching}
\label{subsec:GQ}

Environmental influences on star formation are commonly quantified using the quenched fraction, defined here as $f_{\mathrm{Q}} \equiv N_{\mathrm{quenched}}/N_{\mathrm{total}}$ within a given bin \citep[e.g.][]{Balogh2004, Peng2010, Wetzel2012, Knobel2015, Fossati2017, Davies2019}. Throughout this analysis, galaxies are classified as quenched using the same specific star formation rate (sSFR) criterion adopted in \citetalias{VanKempen2024}: i.e. galaxies with $\log_{10}\mathrm{sSFR}<-11.0~\mathrm{yr^{-1}}$ are considered quenched and all uncertainties and error bars reported throughout are derived via bootstrap resampling with 10{,}000 realisations, and are quoted using the 16th and 84th percentiles.

\begin{figure}[!hbt]
\centering
\includegraphics[width=\linewidth]{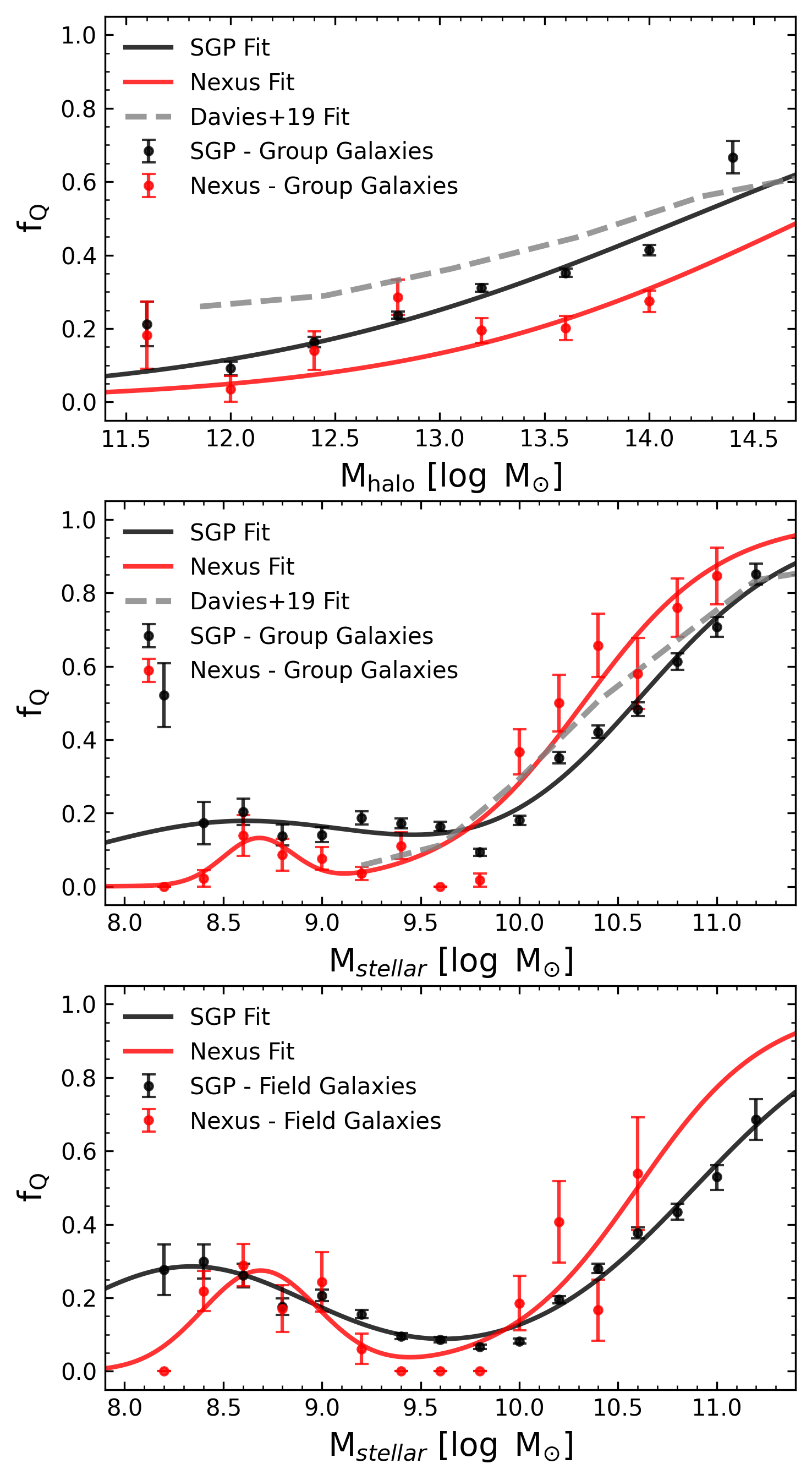}
\caption{Quenched fraction ($f_{\mathrm{Q}}$) as a function of halo mass and stellar mass for group and field galaxies in the SGP (black points) and Nexus (red points). Top panel: $f_{\mathrm{Q}}$ as a function of halo mass for group galaxies. Middle panel: $f_{\mathrm{Q}}$ as a function of stellar mass for group galaxies. Bottom panel: $f_{\mathrm{Q}}$ as a function of stellar mass for field (non-grouped) galaxies. In each panel, points indicate the binned median $f_{\mathrm{Q}}$ values with their associated uncertainties. For the halo mass panel, the best-fitting logistic model (Equation~\ref{eq:logistic}) is shown for the SGP (solid black curve) and Nexus (solid red curve). For the stellar mass panels, the best-fitting Gaussian plus logistic model (Equation~\ref{eq:gauss_logistic}) is shown for the SGP (solid black curve) and Nexus (solid red curve). In the two group galaxy panels (as functions of halo and stellar mass) we overlay the binned values from \citet{Davies2019} as a grey dashed line for comparison.}
\label{fig:Binned_QF}
\end{figure}

Following \citetalias{VanKempen2024}, SFR upper limits are treated in a manner that explicitly accounts for the mass- and distance-dependent sensitivity of the \textit{WISE} W3 band (our main star-formation tracer). Upper limits arise in two distinct regimes: (i) at high stellar masses ($\log M_{\mathrm{stellar}} \geq 10$), where galaxies typically exhibit intrinsically low or negligible star formation, and (ii) at low stellar masses ($\log M_{\mathrm{stellar}} < 10$), where the low surface brightness of galaxies can fall below the W3 detection threshold. For galaxies at $\log M_{\mathrm{stellar}} \geq 10$ and $z < 0.1$, the W3 band is effectively complete for detecting star formation at the levels expected for the star-forming population. In this regime, a SFR upper limit therefore indicates a quenched system, and such galaxies are classified accordingly. However, at lower stellar masses, the detectability of star formation becomes strongly dependent on both surface brightness and distance. For each galaxy with an SFR upper limit in this regime, we therefore evaluate whether its expected W3 flux, given its redshift, lies above or below the survey detection threshold. If the expected flux exceeds the threshold, the upper limit is classified as quenched. Conversely, if the expected flux falls below the detection limit, the upper limit is deemed inconclusive, as the galaxy may host star formation below the sensitivity of the data. In these cases, we conservatively classify the galaxy as star-forming to avoid artificially inflating the quenched fraction.

\begin{table*}
\centering
\caption{Spearman's rank correlation coefficient ($r$) and $p$-value for the relationship between $f_{\mathrm{Q}}$ and either stellar mass ($M_{stellar}$) or halo mass ($M_{\mathrm{halo}}$), for the global SGP and Nexus samples. Best-fit parameters and $1\sigma$ uncertainties for the logistic model (Equation~\ref{eq:logistic}) are reported for the halo mass relations, and for the Gaussian plus logistic model (Equation~\ref{eq:gauss_logistic}) for the stellar mass relations, where $\mu_{\mathrm{g}}$ and $\sigma_{\mathrm{g}}$ are the centre and width of the low-mass Gaussian component, $A_{\mathrm{g}}$ is its amplitude, $M_{50}$ is the characteristic mass at which the logistic component reaches $f_{\mathrm{Q}}=0.5$, and $k$ parametrises the steepness of the high-mass rise.} Results are shown for both group and field environments. 
\label{tab:sgp_nexus_fq_fits}
\resizebox{\textwidth}{!}{%
\begin{tabular}{cccccccc}
\hline
Data & r & p & A$_g$ & $\mu_g$ & $\sigma_g$ & M$_{50}$ [$\log M_{\odot}$] & k \\
\hline
SGP M$_\mathrm{halo}$ & 0.93 & $8.6 \cdot 10^{-4}$ & -- & -- & -- & $14.18 \pm 1.16$ & $0.93 \pm 0.66$ \\
Nexus M$_\mathrm{halo}$ & 0.68 & $9.4 \cdot 10^{-2}$ & -- & -- & -- & $14.75 \pm 1.22$ & $1.07 \pm 0.62$ \\
\hline
SGP (Group) M$_{stellar}$ & 0.55 & $2.6 \cdot 10^{-2}$ & $0.17 \pm 0.03$ & $8.56 \pm 0.43$ & $0.76 \pm 0.50$ & $10.59 \pm 0.05$ & $2.38 \pm 0.37$ \\
Nexus (Group) M$_{stellar}$ & 0.76 & $9.33 \cdot 10^{-4}$ & $0.12 \pm 0.09$ & $8.68 \pm 0.15$ & $0.18 \pm 0.15$ & $10.33 \pm 0.06$ & $2.88 \pm 0.48$ \\
\hline
SGP (Field) M$_{stellar}$ & 0.40 & $1.28 \cdot 10^{-1}$ & $0.28 \pm 0.02$ & $8.33 \pm 0.15$ & $0.63 \pm 0.13$ & $10.89 \pm 0.03$ & $2.26 \pm 0.18$ \\
Nexus (Field) M$_{stellar}$ & 0.21 & $4.93 \cdot 10^{-1}$ & $0.27 \pm 0.09$ & $8.69 \pm 0.11$ & $0.29 \pm 0.11$ & $10.60 \pm 0.13$ & $3.06 \pm 1.17$ \\
\hline
\end{tabular}
}
\end{table*}

\begin{figure*}[t]
\centering
\includegraphics[width=0.98\textwidth]{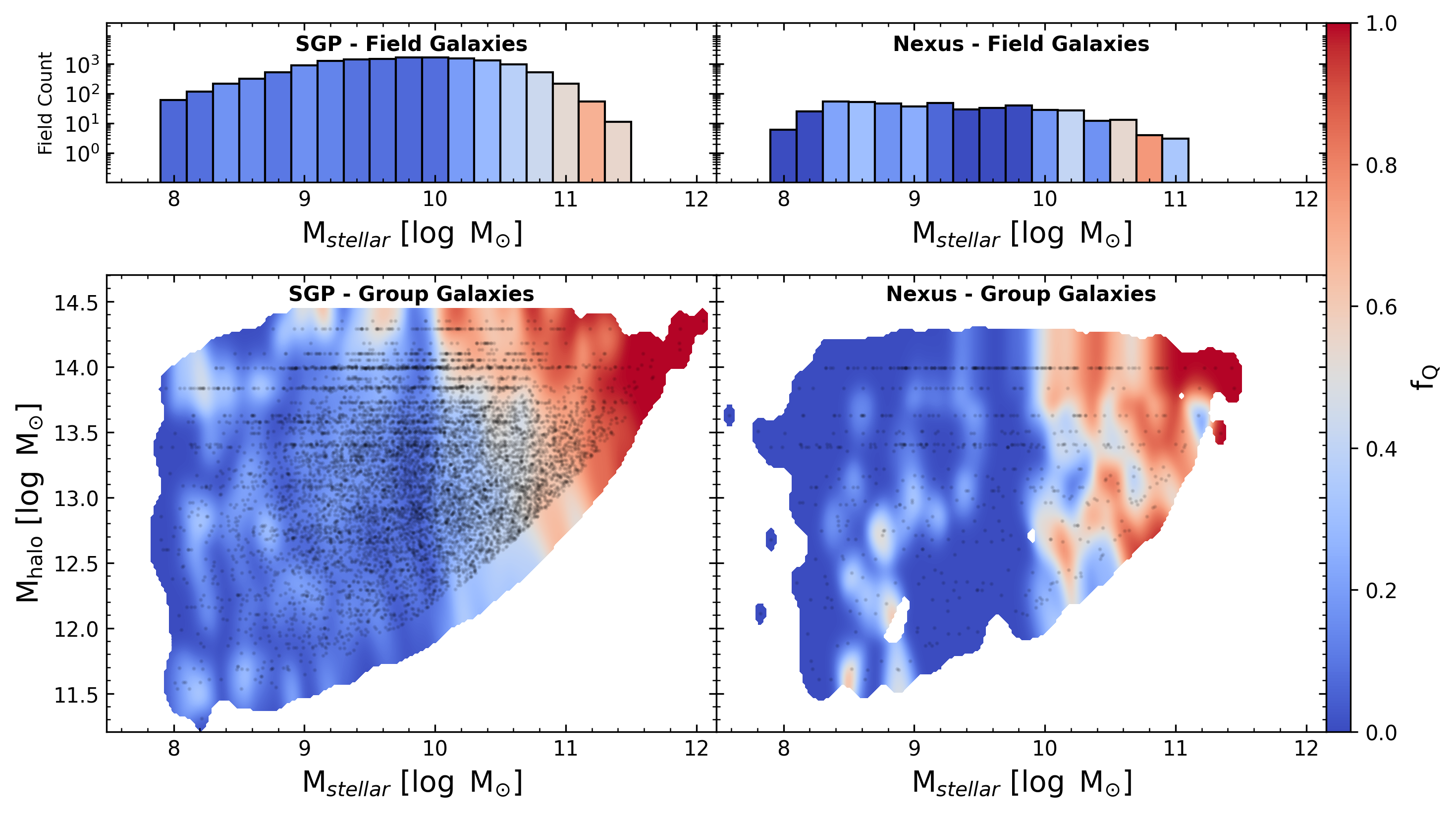}
\caption{
Quenched fraction ($f_{\mathrm{Q}}$), for galaxies in the SGP (left panels) and Nexus (right panels). The top panels show the distribution of field galaxies as a function of stellar mass, colour-coded by the $f_{\mathrm{Q}}$. The bottom panels show $f_{\mathrm{Q}}$ for group galaxies in the stellar--halo mass plane, derived using a smoothed kernel density estimate (KDE), with the underlying galaxy distribution shown by black points. The colour bar indicates the $f_{\mathrm{Q}}$.
}
\label{fig:Global_QF}
\end{figure*}

\begin{figure*}[!hbt]
\centering
\includegraphics[width=\textwidth]{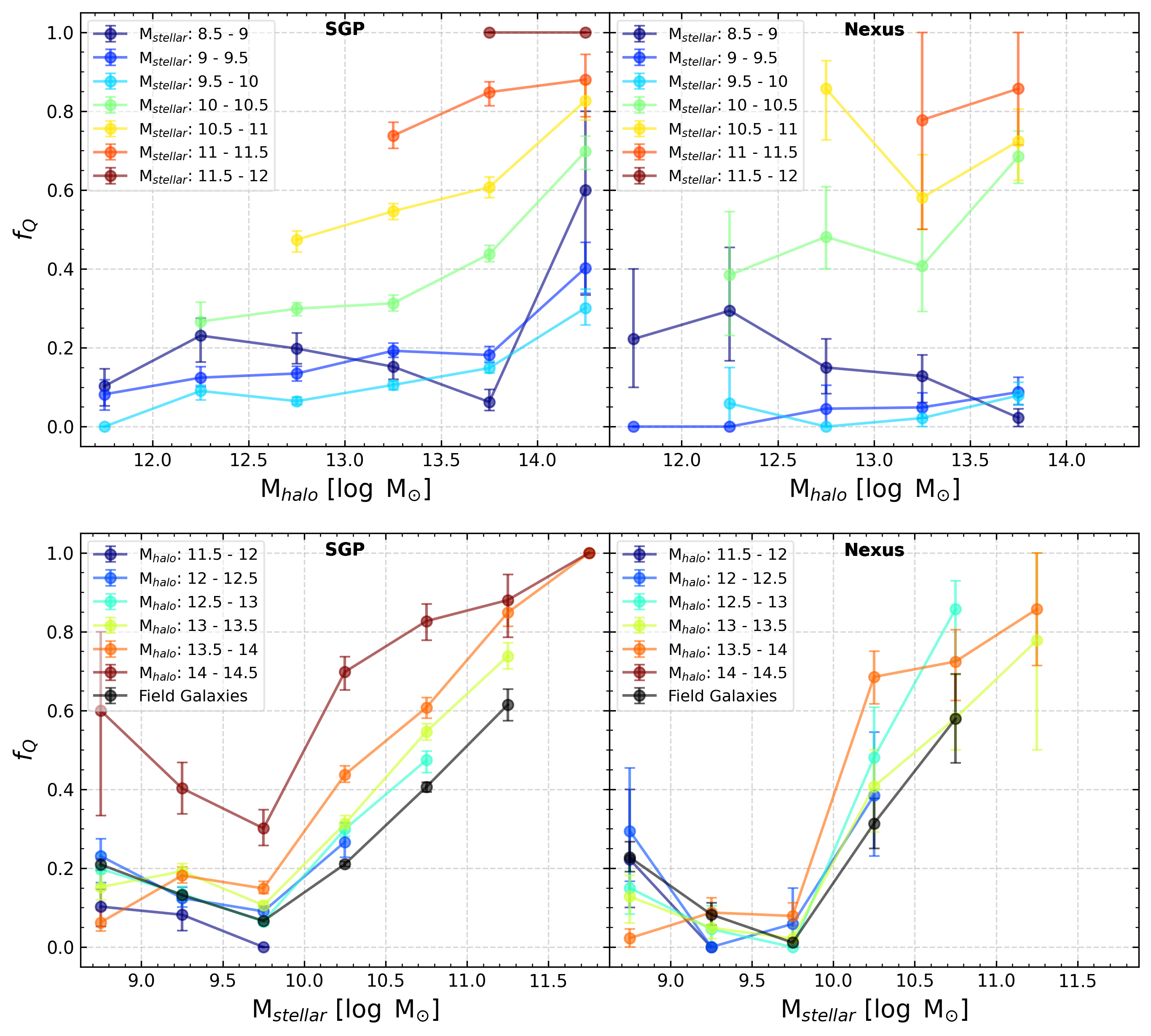}
\caption{Quenched fraction ($f_{\mathrm{Q}}$) for galaxies in the SGP (left panels) and Nexus (right panels). Top panels: $f_{\mathrm{Q}}$ as a function of halo mass (M$_{\mathrm{halo}} \,[\log\, M_\odot]$) shown for different stellar mass bins (M$_{\mathrm{stellar}} \,[\log\, M_\odot]$) for group galaxies. Bottom panels: $f_{\mathrm{Q}}$ as a function of stellar mass for different halo mass bins for group galaxies. In the bottom panels, the $f_{\mathrm{Q}}$ of field galaxies is also shown for comparison. Points indicate binned median values with their associated uncertainties.}
\label{fig:Dissected_QF}
\end{figure*}

Figure~\ref{fig:Binned_QF} presents the quenched fraction, $f_{\mathrm{Q}}$, measured across the SGP and Nexus samples, enabling a direct comparison between an average reference environment and a dynamically assembling superstructure. This comparison provides a baseline against which environmental effects on quenching can be assessed \citep{Balogh2004, Wetzel2012, Fossati2017, vanDerBurg2018}. The figure shows $f_{\mathrm{Q}}$ as a function of halo mass in the top panel, and as a function of stellar mass in the lower two panels, with group galaxies shown in the top and middle panels and field galaxies in the bottom panel. Field galaxies are defined as systems not assigned to any identified group in the adopted SGP group catalogue (see \citetalias{VanKempen2024}). The data are binned in logarithmic mass intervals, with points representing the median $f_{\mathrm{Q}}$ in each bin and associated uncertainties. To ensure statistical robustness while maintaining mass coverage, we include only bins containing at least 10 galaxies in the SGP and at least 5 galaxies in the Nexus. The corresponding binned values are listed in Table~\ref{tab:fq_binned}.

Figure~\ref{fig:Binned_QF} includes a comparison to the group galaxy $f_{\mathrm{Q}}$ values from \citet{Davies2019}, shown as grey dashed lines. The \citet{Davies2019} measurements utilise dynamically-derived halo masses from the GAMA group catalogue \citep{Robotham2011}, providing an independent dataset against which to benchmark our results. The comparison reveals interesting systematic differences between the two approaches, particularly as a function of halo mass. In the halo mass panel, the \citet{Davies2019} results show notably higher $f_{\mathrm{Q}}$ values, especially at group scale halo masses ($\log M_{\mathrm{halo}} < 14$). Conversely, the group stellar mass trends in the middle panel demonstrates a closer agreement between our measurements and those of \citet{Davies2019}, with both datasets exhibiting the similar quantitative features within the mass range probed.

To quantify the overall monotonic trends within these binned relations, we report the Spearman rank correlation coefficients and associated $p$-values in Table~\ref{tab:sgp_nexus_fq_fits}. We emphasise that Spearman's $p$-value measures the strength of a monotonic relationship, and remains a valid and well-defined statistic for nonlinear but monotonically increasing correlations such as those seen in the halo mass and group stellar mass panels. It is only in the presence of genuinely non-monotonic relations, most notably the low-mass bumps visible in the field stellar mass panels, that Spearman's test is expected to underestimate the strength of correlation. In those cases, the low correlation coefficient and high $p$-value should therefore be interpreted as reflecting the non-monotonic character of the relation rather than an absence of correlation, and readers are cautioned against interpreting these values as evidence for a weak or absent trend.

As a function of halo mass, we fit the binned results with a logistic function in the form: \begin{equation} f_{\mathrm{Q}} = \frac{1}{1 + \exp[-k(\log \mathrm{M} - \mathrm{M}_{50})]}, \label{eq:logistic} \end{equation} where $k$ parametrises the transition steepness and $M_{50}$ is the characteristic mass at which $f_{\mathrm{Q}}=0.5$ \citep[e.g.][]{Woo2013, Bluck2016}. As a function of stellar mass, the logistic model provides a poor description of the low-mass end due to the prevalence of a low-mass bump (particularly noticeable with the field population). To better describe this behaviour, we instead fit a Gaussian plus logistic model: \begin{equation} f_{\mathrm{Q}} = A \exp\left[-\frac{(\log M - \mu_{\mathrm{g}})^2}{2\sigma_{\mathrm{g}}^2}\right] + \frac{1}{1 + \exp[-k(\log M - M_{50})]}, \label{eq:gauss_logistic} \end{equation} where the Gaussian component captures the low-mass bump and the logistic component, captures the high mass end. Here, A$_g$ is the amplitude of the low-mass Gaussian bump, $\mu_g$ is the stellar mass at which this excess peaks, and $\sigma_g$ characterises its width in log stellar mass space. The Gaussian amplitude is constrained to be non-negative, ensuring $f_Q \geq 0$ at all masses. Best-fit parameters and $1 \sigma$ uncertainties derived from the covariance matrix are reported in Table~\ref{tab:sgp_nexus_fq_fits}. The corresponding fitted curves are presented in Figure~\ref{fig:Binned_QF} and provide a compact summary for comparing the onset and sharpness of quenching between samples.

The correlation and fit summaries highlight several differences between environments and between the SGP and Nexus. For the group population, $f_{\mathrm{Q}}$ correlates strongly with halo mass in the SGP (Table~\ref{tab:sgp_nexus_fq_fits}), whereas the corresponding halo mass trend in the Nexus is weaker and not statistically significant at the current sample size. We note that this difference is partly driven by the SGP extending to higher halo masses, which broadens the dynamic range of the correlation. In contrast, $f_{\mathrm{Q}}$ shows a clear stellar mass dependence for group galaxies in both regions, with the Gaussian plus logistic fit indicating a lower characteristic transition mass ($M_{50}$) and a steeper transition (larger $k$) in the Nexus group sample relative to the SGP group sample.

For the field population, the Spearman coefficients are low and $p$-values large, but as noted above this reflects the non-monotonic shape of the relation rather than a genuinely weak trend. The Gaussian plus logistic fits reveal a clear low-mass bump followed by a high-mass rise in both the SGP and Nexus field populations, with the high-mass logistic transition occurring at higher $M_{50}$ than for the group population. This is consistent with quenching becoming common only at the high mass end in the absence of identified group-scale environments. We quantify these trends as clearly as possible to facilitate comparison with forthcoming studies, providing a physically motivated decomposition that separately examines the low-mass excess, as well as the onset and sharpness of the high-mass quenching transition, across field and group environments and between the SGP and Nexus.

We note that the SGP samples a broader range of halo masses than the Nexus, extending to higher $M_{\mathrm{halo}}$ and consequently higher $f_{\mathrm{Q}}$ values at the massive end. This difference in dynamic range should be borne in mind when comparing the two samples in the halo mass panel; however, we have verified that this does not materially affect our conclusions. Excluding the highest-mass SGP bins from the fit leaves the best-fit parameters unchanged to within the reported uncertainties, indicating that the elevated SGP correlation is not an artefact of its extended mass baseline but rather reflects a genuinely stronger halo-mass dependence of quenching across the full SGP volume.

In summary, the binned trends of Figure~\ref{fig:Binned_QF}, $f_{\mathrm{Q}}$ increases with stellar mass in both group and field environments in both the SGP and Nexus, although the strength of the relation varies between the two. For group galaxies, $f_{\mathrm{Q}}$ also increases with halo mass in both group and field environments, but with a stronger global halo mass dependence in the SGP than in the Nexus. These one-dimensional measurements provide an accessible summary of how quenching varies with the average mass in each region, but they do not control for the covariance between stellar mass and halo mass.

The scale and richness of our dataset enables us to explore the variation of the $f_{\mathrm{Q}}$ in a two-dimensional plane between halo and stellar mass. Figure~\ref{fig:Global_QF} primarily provides a two-dimensional qualitative view of how the $f_{\mathrm{Q}}$ varies jointly with stellar mass and halo mass, and how this differs between the SGP and the Nexus. The top panels show the stellar mass distributions of field galaxies, with the bar colours indicating $f_{\mathrm{Q}}$ in each stellar mass bin. The bottom panels show group galaxies in the stellar mass and halo mass plane, where the background colour indicates $f_{\mathrm{Q}}$ evaluated on a regular grid. The underlying two-dimensional distribution is estimated using a kernel density estimate (KDE) with adaptive smoothing. The kernel full width at half maximum is chosen to be approximately twice the typical uncertainty in stellar mass (0.15 dex) and halo mass (0.3 dex). This choice is intended to reduce bias from mass measurement scatter in the smoothed density field, and to limit artificial sharpening of $f_{\mathrm{Q}}$ gradients due to underestimated uncertainties. In each grid cell, $f_{\mathrm{Q}}$ is computed as the ratio of the KDE weighted density of quenched galaxies to the total KDE weighted density. Individual galaxies are shown as black points to indicate sampling of the stellar mass and halo mass plane.

As qualitatively demonstrated in Figure~\ref{fig:Global_QF}, we need to control for the covariance between stellar mass and halo mass, Figure~\ref{fig:Dissected_QF} presents the results from Figure~\ref{fig:Global_QF} by binning $f_{\mathrm{Q}}$ trends in controlled bins of the complementary variable. The left columns presents the SGP and the right column shows the Nexus. In the top panels, we show $f_{\mathrm{Q}}$ as a function of halo mass for several bins in stellar mass, and in the bottom panels we present $f_{\mathrm{Q}}$ as a function of stellar mass for a range of halo mass bins. Included in the bottom panel stellar mass panels is the distribution of $f_{\mathrm{Q}}$ for the field galaxies for comparison to non-grouped environments within the SGP and Nexus. For this binned analysis, a minimum of three galaxies per bin was adopted for both SGP and Nexus. This lower threshold was chosen to probe the structure of the trends at fixed stellar mass or fixed halo mass while pushing the current dataset to its practical limits. The trade off is increased statistical noise and a greater sensitivity to sampling variance, most evident in the low and high mass ranges (both stellar and halo) of the sample. The tabulated values corresponding to Figure~\ref{fig:Dissected_QF} are provided in Table~\ref{tab:halo_stellar_fq} and Table~\ref{tab:stellar_halo_fq}.

Together, Figures~\ref{fig:Global_QF} and \ref{fig:Dissected_QF} clarify how the joint stellar mass and halo mass distribution shapes the marginalised trends in Figure~\ref{fig:Binned_QF}. The two-dimensional KDE maps show that high $f_{\mathrm{Q}}$ values are concentrated where both stellar mass and halo mass are large, and they indicate that the SGP and Nexus populate the stellar--halo mass plane differently. Figure~\ref{fig:Dissected_QF} therefore shows explicitly how the apparent dependence on halo mass changes across stellar mass bins, and how the apparent dependence on stellar mass changes across halo mass bins (Table~\ref{tab:halo_stellar_fq} and Table~\ref{tab:stellar_halo_fq}). This demonstrates why controlling for mass is necessary: one-dimensional relations of $f_{\mathrm{Q}}$ versus halo mass (integrating over all stellar masses) or versus stellar mass (integrating over all halo masses) conflate coupled dependencies and can therefore obscure or exaggerate environmental differences between samples \citep[e.g.][]{Peng2012, Wetzel2012, Knobel2015}.

In summary, $f_{\mathrm{Q}}$ increases with both stellar mass and halo mass in the marginalised one-dimensional relations, but Figures~\ref{fig:Global_QF} and \ref{fig:Dissected_QF} show that these dependencies are coupled through the joint distribution of stellar mass and halo mass. Examining $f_{\mathrm{Q}}$ as a function of halo mass within narrow bins of stellar mass, and conversely as a function of stellar mass within narrow bins of halo mass, provides the clearest evidence for how quenching depends jointly on stellar mass and halo mass in the SGP and Nexus.


\subsection{Global Star Formation}
\label{subsec:GSF}

To characterise how star formation is regulated across environment, we measure the mean and scatter of the sSFR distribution for the star-forming population of galaxies in the SGP and Nexus. Throughout this section, we analyse the sSFR ($\log \mathrm{sSFR}\,[\log \mathrm{yr}^{-1}]$) for galaxies classified as star-forming using the same selection and quality cuts described in \citetalias{VanKempen2024}. SFR upper limits are excluded from this analysis; as discussed above, upper limits predominantly arise in systems with intrinsically low star formation activity, particularly at $\log M_{\mathrm{stellar}} \geq 10$, where the WISE W3 band is complete and non-detections robustly indicate quenched galaxies. At lower stellar masses, upper limits can also occur due to surface brightness limitations. However, these cases are conservatively treated as star-forming only when the expected flux falls below the detection threshold. Consequently, the population of galaxies with SFR upper limits is dominated by quenched or weakly star-forming systems, and their removal preferentially excludes the quenched population from the sample. In addition, upper limits cannot be reliably incorporated into binned statistics of $\log \mathrm{sSFR}$. Treating upper limits as detections would bias the mean towards artificially low values, we therefore restrict our analysis to galaxies with robust SFR measurements. Additionally, we require the SFR signal-to-noise ratio to be greater than 2, ensuring that only credible measurements contribute to the derived trends. This selection retains galaxies with sSFR below the canonical star-forming main sequence, including systems with $\log~\mathrm{sSFR} < -11$, thereby capturing galaxies with suppressed, but still detectable, star formation. This approach allows us to probe variations in star formation efficiency without contamination from poorly constrained or non-detected systems. Further details and motivation for these cuts are provided in \citetalias{VanKempen2024}.

\begin{figure}[!hbt]
\centering
\includegraphics[width=\textwidth]{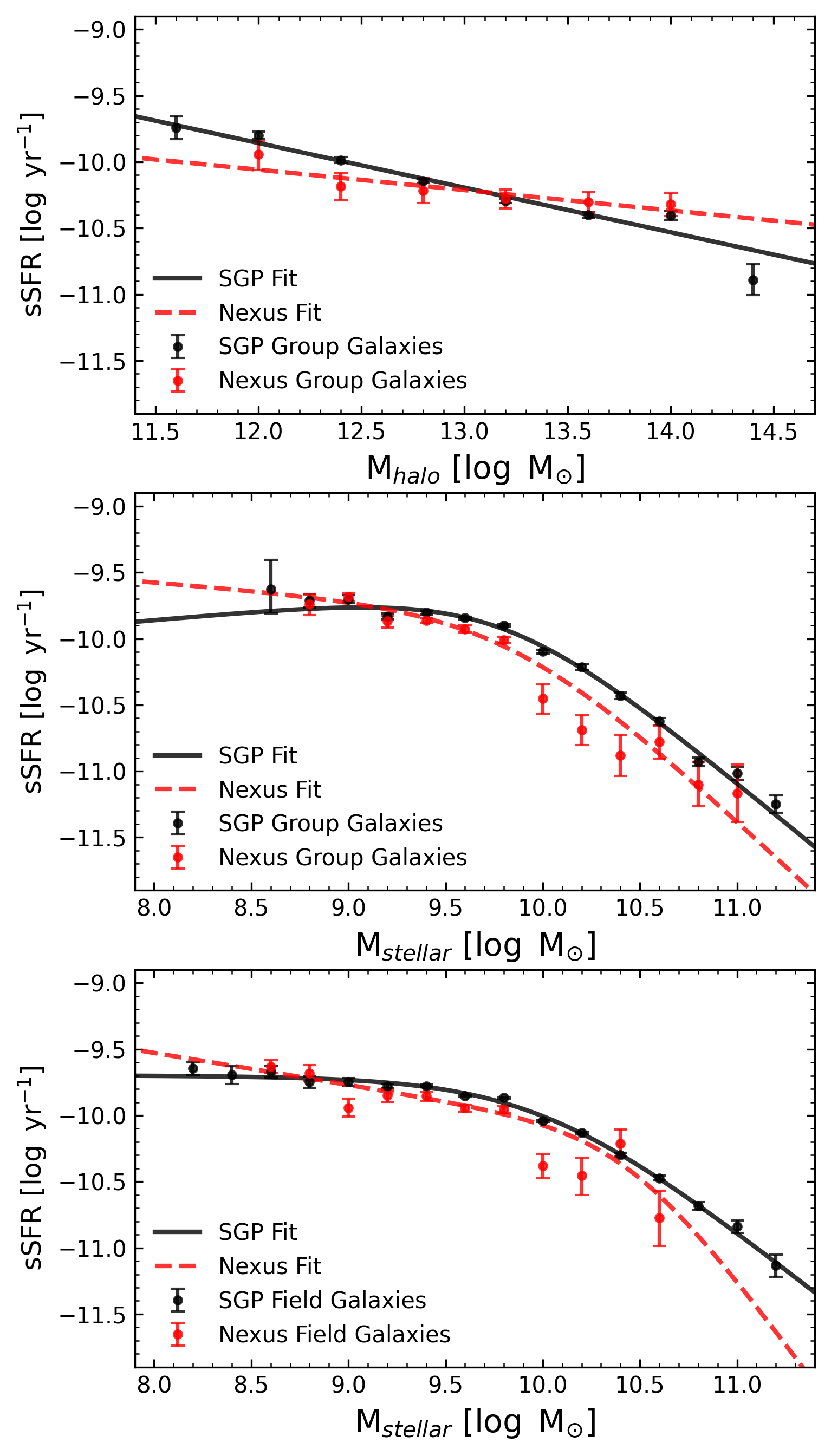}
\caption{Binned mean sSFR (sSFR $[\log \, \mathrm{yr}^{-1}]$) as a function of halo mass and stellar mass for group and field galaxies, with the SGP sample shown as black points and the Nexus sample shown as red points. Top panel: Binned mean sSFR as a function of halo mass for group galaxies. Middle panel: Binned mean sSFR as a function of stellar mass for group galaxies. Bottom panel: Binned mean sSFR as a function of stellar mass for field (non-grouped) galaxies. In each panel, the underlying binned mean sSFR measurements are shown with their associated uncertainties. The best-fitting relations are outlined, with linear fits adopted as a function of halo mass trends a and power-law as a function of stellar mass (see Table~\ref{tab:sgp_nexus_ssfr_fits} for fitting parameters and uncertainties).}

\label{fig:Binned_Mean}
\end{figure}

\begin{table*}
\centering
\caption{Spearman's rank correlation coefficient ($r$) and $p$-value for the relationship between mean $\log \mathrm{sSFR} \, [\log \mathrm{yr}^{-1}]$ and either stellar mass ($M_{stellar}$) or halo mass ($M_{\mathrm{halo}}$), for the global SGP and Nexus samples. The best-fit parameters and their uncertainties to the linear regression for the group galaxies as a function of halo mass and the power-law slope as a function of stellar mass for the group and field galaxies.}
\label{tab:sgp_nexus_ssfr_fits}
\resizebox{\textwidth}{!}{%
\begin{tabular}{cccccccc}
\hline
Data & $r$ & $p$ & m & $M_0$ [$\log M_{\odot}$] & a & b & c \\
\hline
SGP M$_\mathrm{halo}$ & $-1.00$ & $3.11 \cdot 10^{-4}$ & $-0.34\pm0.02$ & -- & -- & -- & $-5.81\pm0.19$ \\
Nexus M$_\mathrm{halo}$ & $-1.00$ & $4.33 \cdot 10^{-3}$ & $-0.15\pm0.06$ & -- & -- & -- & $-8.21\pm0.75$ \\
\hline
SGP (Group) M$_{stellar}$ & $-0.99$ & $6.54 \cdot 10^{-12}$ & -- & $9.80\pm0.09$ & $-10.90\pm0.71$ & $0.13\pm0.08$ & $1.34\pm0.04$ \\
Nexus (Group) M$_{stellar}$ & $-0.98$ & $3.09 \cdot 10^{-8}$ & -- & $9.86\pm0.31$ & $-8.61\pm1.62$ & $-0.12\pm0.18$ & $1.25\pm0.13$  \\
\hline
SGP (Field) M$_{stellar}$ & $-0.99$ & $1.09 \cdot 10^{-13}$ & -- & $10.01\pm0.07$ & $-9.66\pm0.31$ & $-0.01\pm0.04$ & $1.16\pm0.04$ \\
Nexus (Field) M$_{stellar}$ & $-0.92$ & $6.61 \cdot 10^{-5}$ & -- & $10.45\pm0.16$ & $-7.60\pm0.66$ & $-0.24\pm0.07$ & $1.76\pm1.22$ \\
\hline
\end{tabular}
}
\end{table*}

\begin{figure*}[!hbt]
\centering
\includegraphics[width=\textwidth]{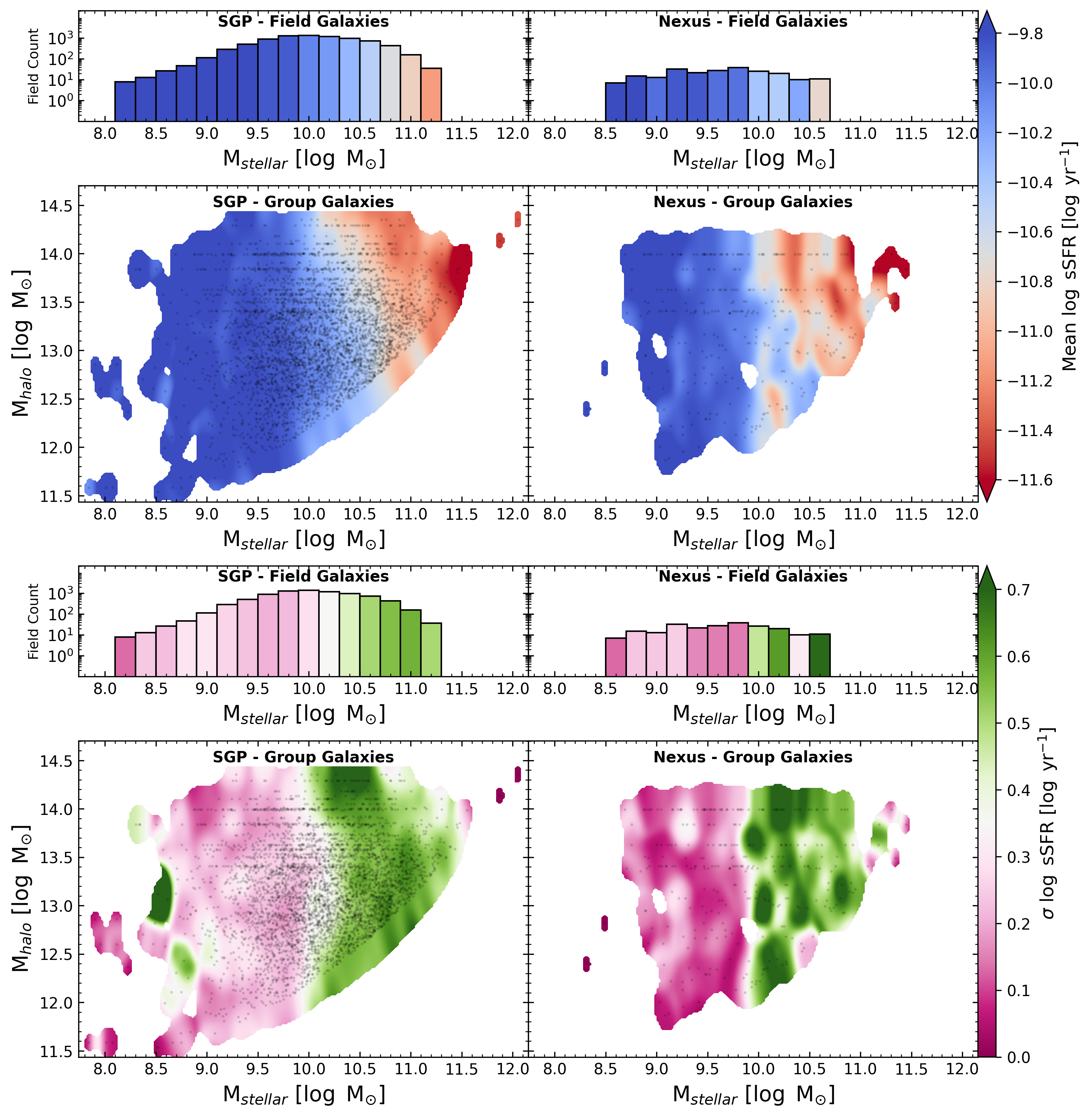}
\caption{
Mean and scatter of sSFR in the stellar--halo mass plane for group and field galaxies in the SGP (left panels) and Nexus (right panels). The top two panels show the mean sSFR: the top histogram panels display the mean sSFR of field galaxies as a function of stellar mass, while the second row shows the mean sSFR of group galaxies as a function of stellar and halo mass, derived using a smoothed KDE. The bottom two panels show the scatter in sSFR, quantified as the standard deviation ($\sigma \log \mathrm{sSFR}\,[\log\,\mathrm{yr}^{-1}]$). The bottom histogram panels show the sSFR scatter for field galaxies as a function of stellar mass, while the bottom row shows the sSFR scatter for group galaxies in the stellar--halo mass plane, also estimated using a smoothed KDE. In the KDE panels, the underlying galaxy distributions are indicated by black points.
}
\label{fig:Global_SF}
\end{figure*}

\begin{figure*}[!hbt]
\centering
\includegraphics[width=\textwidth]{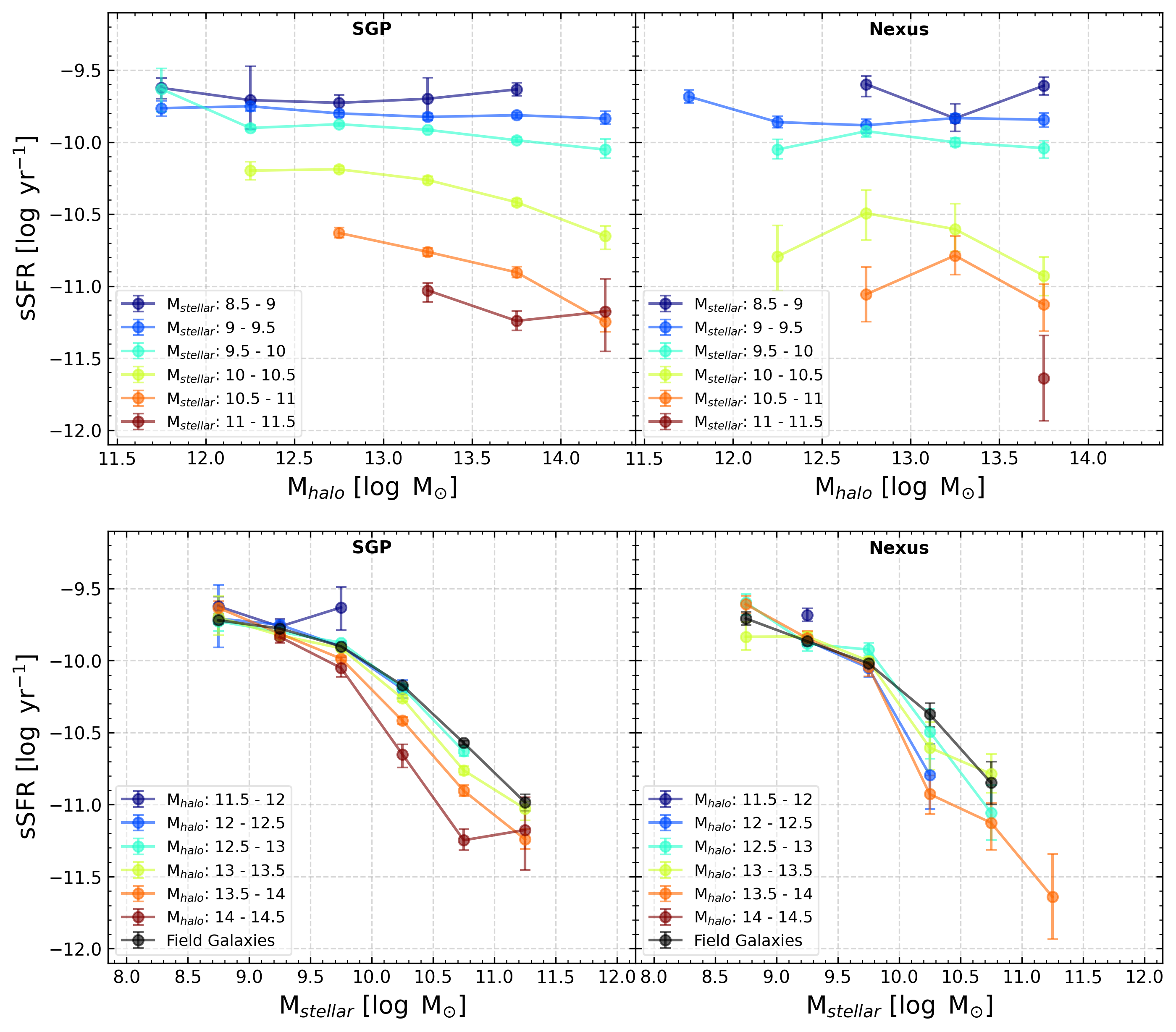}
\caption{Binned mean sSFR (sSFR $[\log \, yr^{-1}]$) for galaxies in the SGP (left panels) and Nexus (right panels). Top panels: Binned mean sSFR fraction as a function of halo mass (M$_{\mathrm{halo}} \,[\log\, M_\odot]$) shown for different stellar mass bins (M$_{\mathrm{stellar}} \,[\log\, M_\odot]$) for group galaxies. Bottom panels: Binned mean sSFR as a function of stellar mass for different halo mass bins for group galaxies. In the bottom panels, the binned mean sSFR of field galaxies is also shown for comparison. Points indicate binned median values with their associated uncertainties.}
\label{fig:Dissected_Mean}
\end{figure*}


\begin{figure*}[!hbt]
\centering
\includegraphics[width=\linewidth]{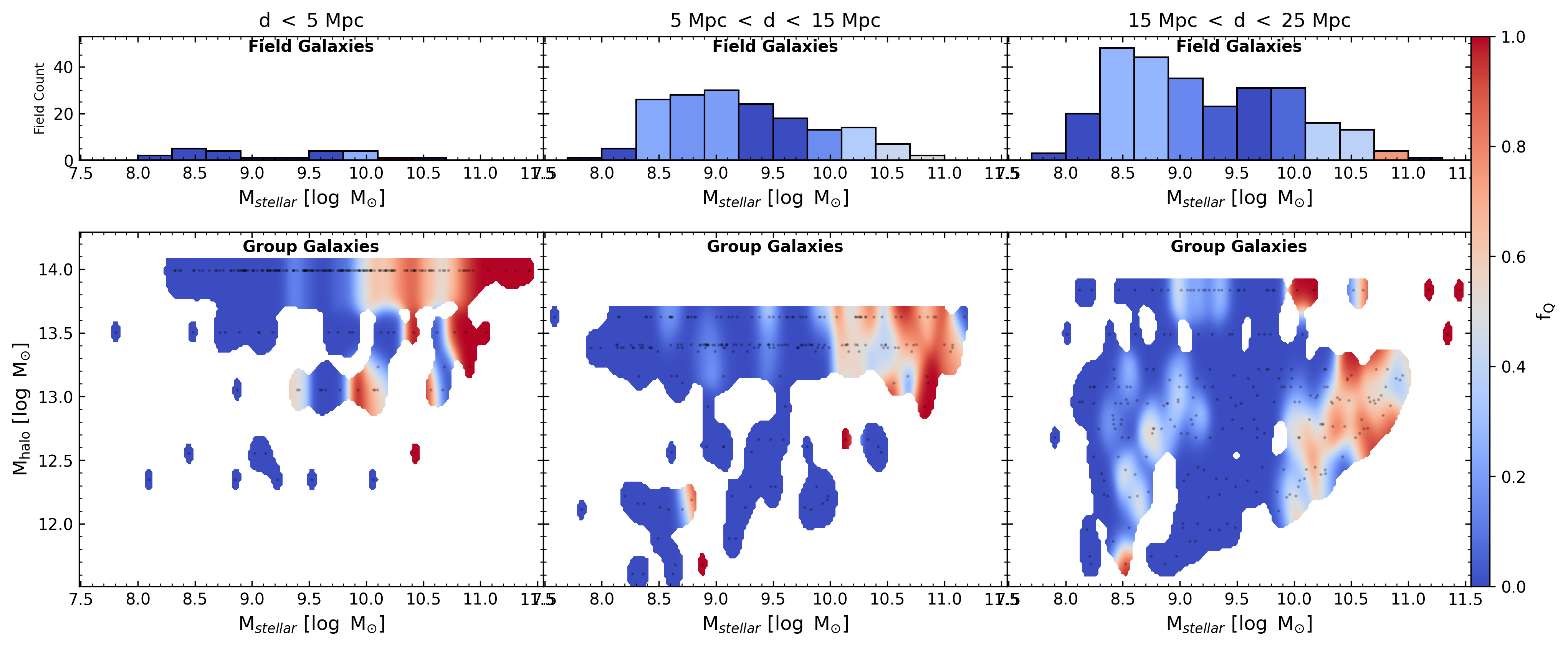}
\caption{
Distribution of the Quenched fraction ($f_{\mathrm{Q}}$) in the stellar mass–halo mass plane for galaxies in the Nexus region. The left panels show galaxies within the 5 Mpc shell from A4038, the middle panels show galaxies between 5 and 15 Mpc, and the right panels show galaxies between 15 and 25 Mpc. The bottom panels indicate the colour-coded $f_{\mathrm{Q}}$ for grouped galaxies as a function of both stellar mass ($\log~M_{stellar}~[M_{\odot}]$) and halo mass ($\log~M_{halo}~[M_{\odot}]$), with the colour scale ranging from blue (predominantly star-forming) to red (predominantly quenched). The underlying distribution is computed using a two-dimensional KDE (KDE) with adaptive smoothing, where the $f_{\mathrm{Q}}$ represents the ratio of the KDE-weighted density of quenched galaxies to the total KDE-weighted density in each cell. Black points indicate the positions of individual galaxies within the sample. The top panels presents a coloured histogram showing the $f_{\mathrm{Q}}$ distribution of field galaxies as a function of $\log~M_{stellar}~[M_{\odot}]$.
}
\label{fig:qf_zones}
\end{figure*}

\begin{figure*}[!hbt]
\centering
\includegraphics[width=\linewidth]{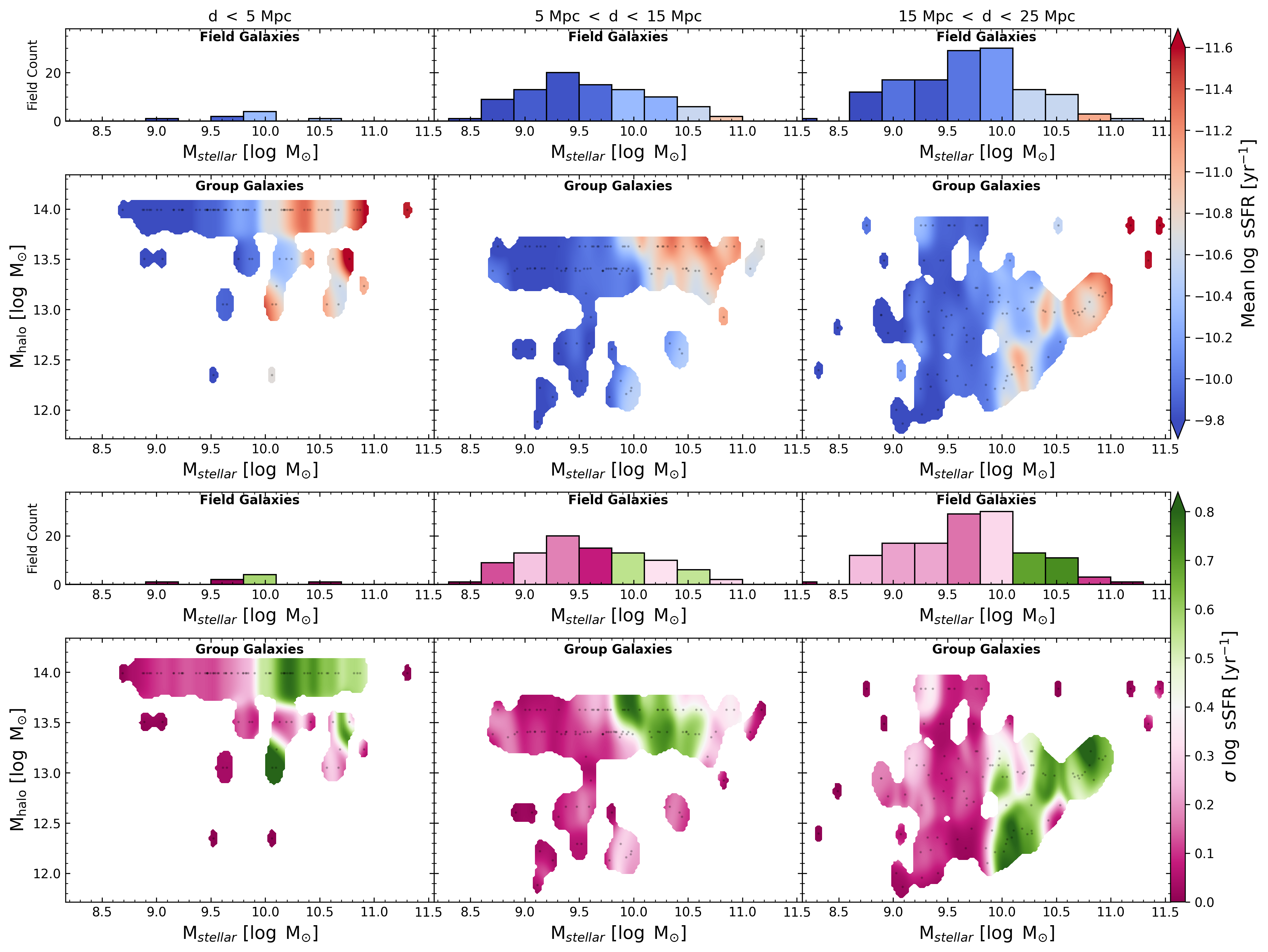}
\caption{
Mean (top row) and RMS scatter (bottom row) of $\log{\rm sSFR}$ in the stellar–halo mass plane for Nexus group galaxies, split by cluster-centric distance from A4038. Columns show $d<5$~Mpc (left), $5<d<15$~Mpc (middle), and $15<d<25$~Mpc (right). In each panel, colours are computed with a two-dimensional adaptive KDE over $\log M_{stellar}$ and $\log M_{\rm halo}$; black points mark individual group galaxies and cells with insufficient sampling are masked (white). Top row: colour-coded mean $\log{\rm sSFR}$ [yr$\pm^{-1}$], from blue (higher sSFR) to red (lower sSFR). Bottom row: colour-coded RMS of $\log{\rm sSFR}$, from magenta (lower scatter, more uniform sSFR) to green (higher scatter, more varied sSFR). The histograms above each column show the field-galaxy counts as a function of $\log M_{stellar}$ for the corresponding radial shell, providing a baseline independent of group environment.}
\label{fig:ssfr_zones}
\end{figure*}

Figure~\ref{fig:Binned_Mean} presents the binned median of the mean $\log \mathrm{sSFR}$ against logarithmic mass bins for the SGP (black points) and the Nexus (red points) respectively. The top row shows the mean $\log \mathrm{sSFR}$ of star-forming group galaxies as a function of halo mass ($M_{\mathrm{halo}}$), while the middle and bottom rows present the mean $\log \mathrm{sSFR}$ as a function of stellar mass ($M_\mathrm{stellar}$) for star-forming group and field galaxies, respectively. In each panel, points show the binned median of the mean values, with associated uncertainties. For these global one-dimensional trends, only bins containing more than eight galaxies in the SGP and more than four galaxies in the Nexus are considered, to reduce sensitivity to small number statistics whilst attempting to probe the extremes of our mass ranges for the large SGP sample and the smaller Nexus sample. The binned values underlying Figure~\ref{fig:Binned_Mean} are reported in Table~\ref{tab:ssfr_binned}.

To quantify the overall trends in the binned relations, Spearman rank correlation coefficients and associated $p$ values are reported in Table~\ref{tab:sgp_nexus_ssfr_fits}. For trends as a function of halo mass, a linear relation of the form $\log\mathrm{sSFR} = m\,\log M_{\mathrm{halo}} + c$ is fitted to the binned points. For trends as a function of stellar mass, we instead adopt a power-law motivated fit that allows for curvature in the $\log \mathrm{sSFR}$--$\log M_{\star}$ relation, thereby better capturing the flattening of the star-forming sequence at low stellar masses \citep[e.g.][]{Noeske2007, Whitaker2012, Speagle2014, Schreiber2015}. This fit is in the form: \begin{equation} \log_{10}\mathrm{sSFR} = a + b\log_{10}M_{\mathrm{stellar}} - \log_{10}\left[1 + \left(\frac{M_{\mathrm{stellar}}}{M_{0}}\right)^{c}\right], \label{eq:ssfr_powerlaw} \end{equation} where a is a normalisation constant, b is the low-mass power-law slope, $M_{0}$ is the characteristic turnover mass above which the sSFR begins to decline, and c controls the sharpness of that turnover.

The correlation and fit summaries provide a compact comparison of how strongly star formation is regulated by stellar mass and halo mass, and whether this differs significantly between the SGP and the Nexus. The mean $\log \mathrm{sSFR}$ shows very strong anti-correlations with stellar mass for both group and field galaxies in both regions (Table~\ref{tab:sgp_nexus_ssfr_fits}), with $r \leq -0.98$ in all cases except for the Nexus field sample ($r=-0.92$), and all correlations remaining highly significant ($p \ll 0.01$). The fitted power-law parameters indicate that the decline of $\log \mathrm{sSFR}$ with stellar mass is steeper for group galaxies than for field galaxies. In the SGP, the power-law slope parameter is $a=-10.90\pm0.71$ for group galaxies compared to $a=-9.66\pm0.31$ for field galaxies, while in the Nexus the corresponding values are $a=-8.61\pm1.62$ (group) and $a=-7.60\pm0.66$ (field), albeit with larger uncertainties.

For group galaxies, the binned mean $\log \mathrm{sSFR}$ also declines with halo mass in both regions, with Spearman coefficients of $r=-1.00$ in both the SGP and Nexus samples. The fitted linear slopes indicate a stronger halo mass dependence in the SGP ($m=-0.34\pm0.02$) compared to the Nexus ($m=-0.15\pm0.06$), although the latter is subject to larger uncertainties due to the smaller sample size. Interpreting these coefficients requires caution because the fitted relations are summary statistics of binned trends; however, they provide a uniform basis for comparing the relative steepness of the mean relations between stellar mass and halo mass, and between the SGP and Nexus environments.

Figure~\ref{fig:Global_SF} provides a qualitative two-dimensional view of how star formation varies in the stellar--halo mass plane for group galaxies, complemented by stellar mass histograms for the field population. The top pair of panels presents a KDE smoothed mean sSFR (mean $\log \mathrm{sSFR}$) as a function of halo mass and stellar mass, whilst the bottom pair of panels shows the one standard deviation scatter of the KDE smoothed mean sSFR value ($\sigma~\log~\mathrm{sSFR}$). The analysis of the KDE smoothed mean and one standard deviation maps use the same methodology as in Section~\ref{subsec:GQ}. For the mean map, each grid cell is computed as the KDE weighted mean of $\log \mathrm{sSFR}$ for star-forming galaxies. For the standard deviation map, $\sigma$, each grid cell is calculated from KDE weighted moments of the $\log \mathrm{sSFR}$ distribution, by combining the KDE weighted mean of $\log \mathrm{sSFR}$ with the KDE weighted mean of $(\log \mathrm{sSFR})^{2}$ to recover the cell variance and hence the standard deviation. These maps are intended as qualitative diagnostics of joint trends and of regions where the star-forming population exhibits elevated intrinsic diversity.

To control for the covariance between stellar mass and halo mass in their relation with sSFR, as demonstrated in Figure~\ref{fig:Global_SF}, Figure~\ref{fig:Dissected_Mean} presents the mean $\log \mathrm{sSFR}$ binned by halo mass at fixed stellar mass, and by stellar mass at fixed halo mass. The left column shows SGP and the right column shows Nexus. The top panels show mean $\log \mathrm{sSFR}$ as a function of halo mass within stellar mass bins. The bottom panels show mean $\log \mathrm{sSFR}$ as a function of stellar mass within halo mass bins, with the field relation included for reference. For this analysis, a minimum of three galaxies per bin is adopted for both the SGP and the Nexus, with bins below this threshold excluded. The corresponding binned values are reported in Table~\ref{tab:halo_stellar_sSFR} and Table~\ref{tab:stellar_halo_sSFR}.

Taken together, Figures~\ref{fig:Global_SF} and \ref{fig:Dissected_Mean} show that the aggregate halo mass trend in Figure~\ref{fig:Binned_Mean} is not uniform across stellar mass. When the sample is split into stellar mass bins (Figure~\ref{fig:Dissected_Mean}, top panels), the halo mass dependence varies in strength and, in some bins, becomes comparatively weak within the uncertainties. Conversely, when split by halo mass and examined as a function of stellar mass (Figure~\ref{fig:Dissected_Mean}, bottom panels), the decline of mean $\log \mathrm{sSFR}$ with stellar mass remains evident across halo mass bins, indicating that the dominant driver of the mean trend is stellar mass, with halo mass providing an additional modulation that can be masked or amplified when the sample is marginalised. The two-dimensional mean maps in Figure~\ref{fig:Global_SF} are consistent with this coupled behaviour, showing structure across the joint stellar mass and halo mass plane rather than a single separable dependence. The standard deviation maps add complementary information by showing that $\sigma~\log~\mathrm{sSFR}$ varies more strongly with stellar mass than with halo mass across the parameter space occupied by the star-forming group population.

As shown prior, controlling for mass is necessary when interpreting the one-dimensional binned trends in Figure~\ref{fig:Binned_Mean}. A relation of mean $\log \mathrm{sSFR}$ versus halo mass that integrates over all stellar masses mixes galaxies with systematically different typical sSFR, and a relation of mean $\log \mathrm{sSFR}$ versus stellar mass that integrates over all halo masses mixes environments with different levels of star formation regulation. The controlled comparisons in Figure~\ref{fig:Dissected_Mean}, together with the joint view in Figure~\ref{fig:Global_SF}, therefore provide the clearest description of how mean star-forming activity and its diversity depend jointly on stellar mass and halo mass in the SGP and Nexus.

Overall, the global diagnostics provide a consistent empirical picture of star-forming activity across the two regions. The one-dimensional binned relations show that mean $\log \mathrm{sSFR}$ declines with increasing stellar mass in both group and field environments, and that an average decline with halo mass is also present for group galaxies. The controlled binning and two-dimensional maps show that these dependencies are coupled through the joint distribution of stellar mass and halo mass, with stellar mass setting the dominant decline of mean $\log \mathrm{sSFR}$ and halo mass acting as an additional modulation that varies across stellar mass. The standard deviation maps further indicate that the diversity of star-forming activity varies more strongly with stellar mass than with halo mass over the range probed here, and that the SGP measurements provide the higher signal to noise baseline against which the corresponding Nexus patterns can be compared.


\subsection{Radial Trends in the Nexus}
\label{subsec:Nex_Radial}

The Nexus region provides a direct opportunity to examine how quenching and star formation vary with distance from the cosmic node and forming supercluster with A4038 at the centre. Motivated by the expectation that environmental processing changes systematically with local density, we analyse both the $f_{\mathrm{Q}}$ and the star-forming population properties as a function of stellar mass and halo mass within the three radial zones defined in Section~\ref{subsec:nexus_zones}. These zones span the inner core (0 to 5 Mpc), an intermediate region (5 to 15 Mpc), and an outer region (15 to 25 Mpc). Given the construction described in Section~\ref{subsec:nexus_zones}, the zone assignments are expected to be highly robust for the purposes of this qualitative comparison, and we explicitly compare zones to highlight how the observed patterns vary with radius.

Figure~\ref{fig:qf_zones} shows the distribution of the $f_{\mathrm{Q}}$ within each radial zone. Each column corresponds to one zone, with the field population shown as a stellar mass histogram in the top row and the group population shown in the stellar mass and halo mass plane in the bottom row. In each zone, $f_\mathrm{Q}$ is estimated using the same two-dimensional KDE approach adopted in Section~\ref{subsec:GQ}, applied independently to the galaxies in that zone, with the background colour indicating the KDE based $f_\mathrm{Q}$ and black points indicating the underlying galaxy distribution.

Figure~\ref{fig:ssfr_zones} presents the corresponding radial trends for the star-forming population, using the same Nexus zone definitions. For this analysis we use the same star-forming sample defined in Section~\ref{subsec:GSF}, following the selection described in \citetalias{VanKempen2024}. The figure is arranged by column, with one column per zone, and it shows both the mean $\log \mathrm{sSFR}\,[\log \mathrm{yr}^{-1}]$ and the standard deviation, $\sigma$, of $\log \mathrm{sSFR}$. 

The $f_\mathrm{Q}$ and mean $\log \mathrm{sSFR}$ maps in Figures \ref{fig:qf_zones} and \ref{fig:ssfr_zones} show clear structure across the plane in all three zones, with lower $f_\mathrm{Q}$ and mean $\log \mathrm{sSFR}$ values occurring preferentially towards higher stellar mass, and additional variation with halo mass at fixed stellar mass in parts of the plane. Differences between zones are evident in the range of stellar mass and halo mass that is populated and in the spatial extent of low $f_\mathrm{Q}$ and mean $\log \mathrm{sSFR}$ regions ($f_\mathrm{Q} < 0.6$, $\log \mathrm{sSFR} < -10.8$), with the inner zone showing the strongest masking and small scale incompleteness due to limited sampling. The standard deviation maps show a distinct qualitative behaviour, where the strongest gradients in $\sigma$ occur along the stellar mass direction, with comparatively weaker variation with halo mass at fixed stellar mass. The field histograms provide additional context by showing how the stellar mass distribution of star-forming field galaxies changes across the three zones, and by indicating that, at high stellar mass, field galaxies remain less quenched and more strongly star-forming than group galaxies at similar stellar masses, while the stellar mass dependence of $\sigma$ is broadly similar between the field and group populations over the stellar mass range where the field statistics are well constrained.

Apparent differences between zones are present in the group population, but these should not be interpreted as evidence for an intrinsic distance-driven trend. The occupied regions of the stellar mass--halo mass plane differ strongly between zones, with each zone sampling a largely distinct halo-mass range and only a small number of groups. Accordingly, the observed changes in $f_{\mathrm{Q}}$ and mean/$\sigma$ sSFR across zones can be explained by differences in the halo mass function probed in each zone, rather than by projected distance from A4038 alone. The field population provides complementary context, but the same sampling limitation applies when comparing zone-to-zone behaviour.

Taken together, Figures~\ref{fig:qf_zones} and \ref{fig:ssfr_zones} demonstrate that the sampled stellar mass--halo mass domain differs between zones, and that this difference strongly conditions the inferred $f_\mathrm{Q}$ and sSFR patterns. We therefore do not infer that zone-to-zone variation is driven by projected distance itself. With the current data, the apparent radial differences are plausibly dominated by the differing halo-mass ranges sampled in each zone, compounded by small-number statistics. Robustly testing whether there is an additional large-scale environmental dependence at fixed stellar mass and fixed halo mass will require combining multiple analogous structures and analysing substantially larger survey volumes; importantly, even the SGP reference volume remains limited in its sampling of the most massive halos.


\subsection{A4038 Projected Phase–Space}
\label{subsec:PPS}

\begin{figure}[!hbt]
\centering
\includegraphics[width=\linewidth]{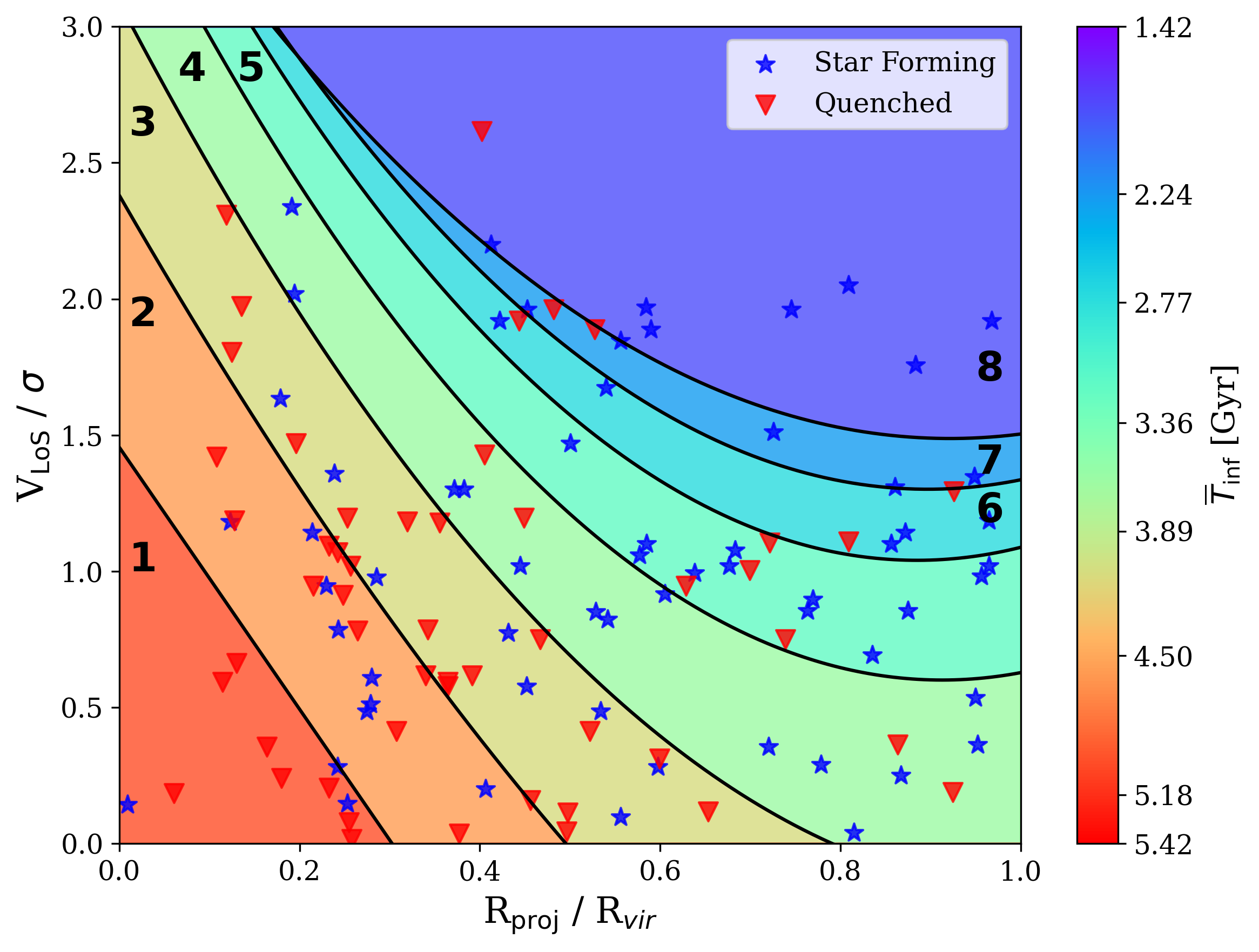}
\caption{Phase-space diagram of A4038, showing eight infall zones based on mean infall time, adapted from \cite{Pasquali2019}. The x-axis represents the projected cluster-centric radius normalised to the virial radius, and the y-axis shows the absolute line-of-sight velocity normalised to the velocity dispersion. The background colour corresponds to the mean infall time in gigayears, as indicated by the colour bar. Blue stars represent star-forming galaxies, while red triangles indicate quenching galaxies.}
\label{fig:Phase_Space}
\end{figure}

As a final, cluster-scale view that complements the Nexus-wide radial trends (Section~\ref{subsec:Nex_Radial}), we use the projected phase--space (PPS) diagram of A4038 to relate quenching to orbital phase within a single system. We consider PPS coordinates of the ratio between the projected radius of cluster members and the virial radius of the cluster ($R_{\rm proj}/R_{\rm vir}$) and the absolute line-of-sight velocity between the cluster and the cluster galaxies normalised by the cluster velocity dispersion ($|V_{\rm LOS}|/\sigma$), and assign galaxies to eight PPS regions following the calibration of \citet{Pasquali2019}, which used the Yonsei Zoom-in Cluster Simulations \citep[YZiCS;][]{Choi2017} to map positions within the PPS to a mean infall time, $\overline{T}_{\rm inf}$. The adopted zone boundaries and corresponding $\overline{T}_{\rm inf}$ values are shown in Figure~\ref{fig:Phase_Space}. In this scheme, zones 1--3 correspond to earlier infall (older typical $\overline{T}_{\rm inf}\!\gtrsim\!4.5$~Gyr) and zones 6--8 correspond to recent or first infall (younger typical $\overline{T}_{\rm inf}\!\lesssim\!2.8$~Gyr).

\begin{table}
\centering
\caption{Mean PPS infall-times from \citet{Pasquali2019} and adopted cluster phases. Values are means with asymmetric uncertainties given by the 16th and 84th percentiles.}
\label{tab:mean_infalls}
\begin{tabular}{ccc}
\hline
Zone & $\overline{T}_{\rm inf}$ [Gyr] & Phase \\
\hline
1 & $5.42^{+1.71}_{-2.01}$ & \multirow{2}{*}{Late-Phase} \\
2 & $5.18^{+1.95}_{-2.16}$ \\
\hline
3 & $4.50^{+1.89}_{-2.17}$ & \multirow{3}{*}{Mid-Phase} \\
4 & $3.89^{+1.77}_{-1.79}$ \\
5 & $3.36^{+1.72}_{-1.78}$\\
\hline
6 & $2.77^{+1.43}_{-1.78}$ & \multirow{3}{*}{Early-Phase} \\
7 & $2.24^{+0.47}_{-1.46}$\\
8 & $1.42^{+0.38}_{-0.93}$\\
\hline
\end{tabular}
\end{table}

\begin{figure}[!hbt]
\centering
\includegraphics[width=\linewidth]{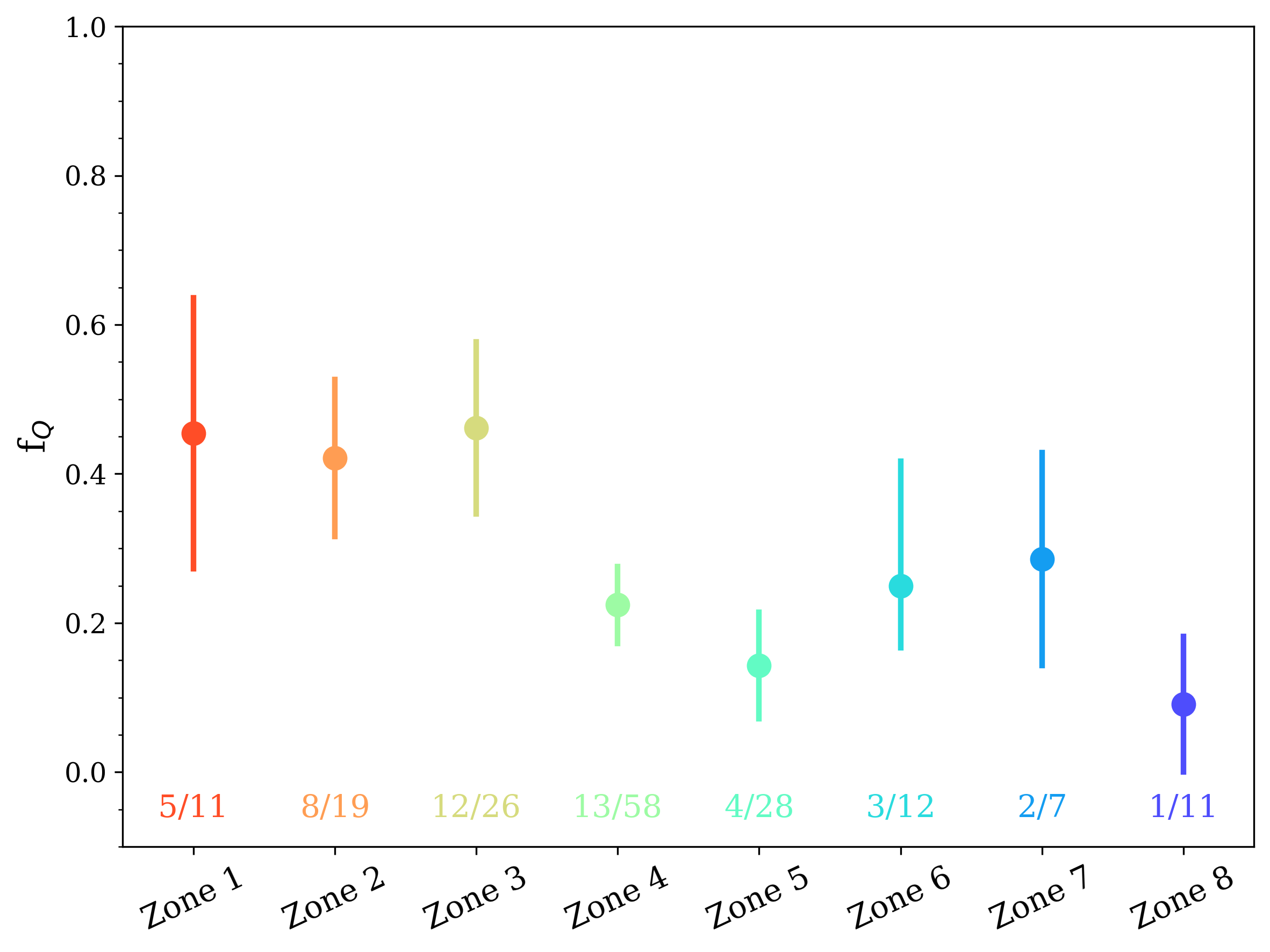}
\includegraphics[width=\linewidth]{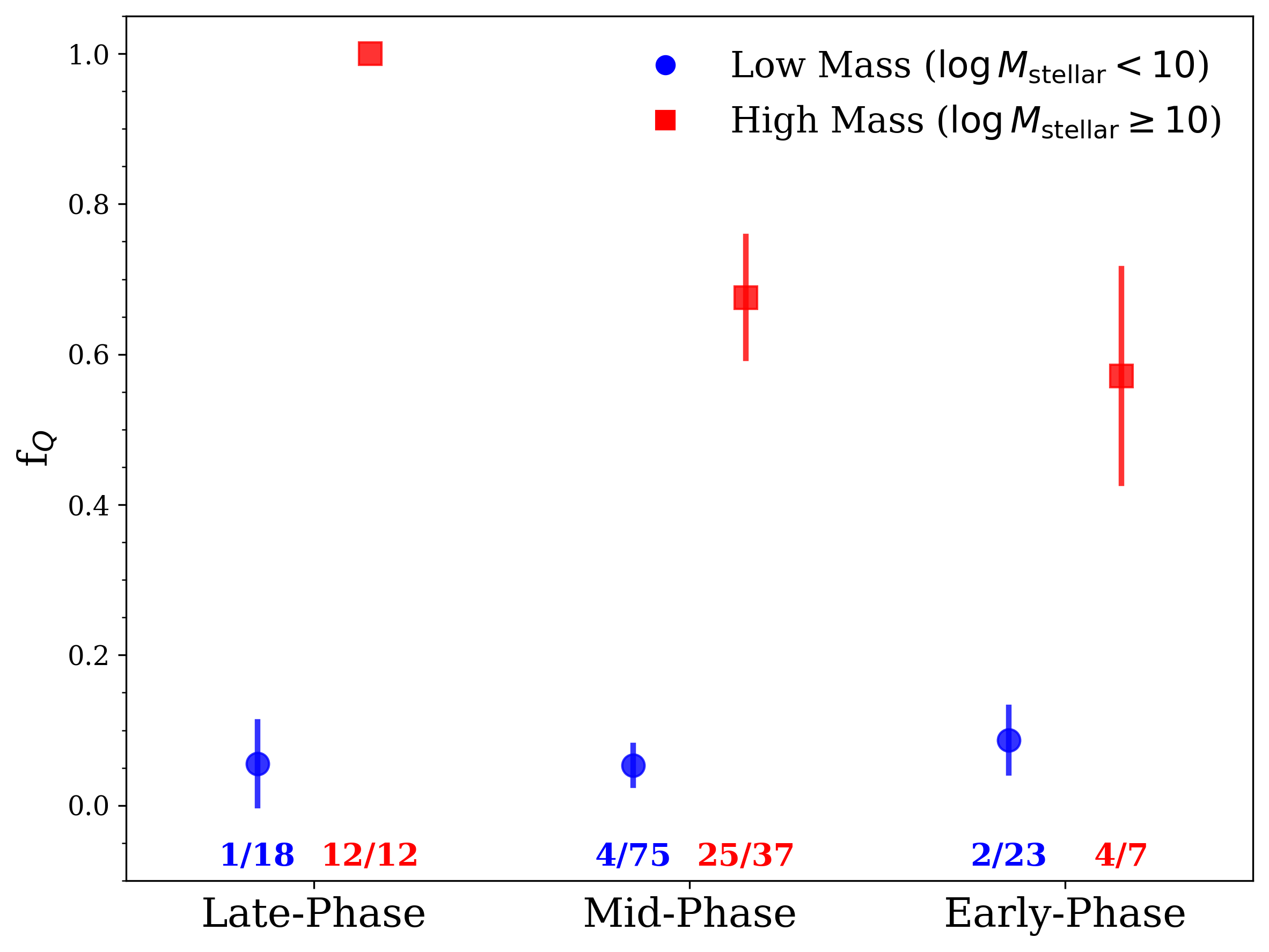}
\caption{Quenched fraction as a function of mean infall time for galaxies in A4038. Top panel: $f_{\rm Q}$ as a function of the eight infall zones defined in Figure \ref{fig:Phase_Space}. Bottom panel: $f_{\rm Q}$ grouped by infall phase (Late, Mid, and Early), further subdivided into low-mass ($\log M_{\rm stellar} < 10$; blue circles) and high-mass ($\log M_{\rm stellar} \geq 10$; red squares) populations to illustrate mass-dependent quenching. In both panels, error bars follow the global uncertainty prescription introduced at the start of this section. The fractions indicated below indicate the number of quenched galaxies relative to the total number of galaxies in that specific bin.}
\label{fig:QF_Phase_Space}
\end{figure}

The $f_{\mathrm{Q}}$ is measured independently within each PPS zone (Figure~\ref{fig:QF_Phase_Space}) using the same quenched classification adopted throughout this analysis. In the top panel, $f_{\mathrm{Q}}$ is calculated for each of the eight PPS zones as the fraction of quenched galaxies relative to the total population, with associated uncertainties. In the bottom panel, we extend this analysis to account for stellar mass dependence shown in Sections~\ref{subsec:GQ} and \ref{subsec:GSF} while preserving sufficient number statistics by grouping the PPS zones into three infall phases (Late, Mid, and Early), and computing $f_{\mathrm{Q}}$ separately for low-mass ($\log M_{\rm stellar}<10$) and high-mass ($\log M_{\rm stellar}\geq10$) galaxies. The mean infall times for each zone, along with the 16th--84th percentile ranges that capture projection effects and orbital diversity, are listed in Table~\ref{tab:mean_infalls}.

The top panel of Figure~\ref{fig:QF_Phase_Space} shows that $f_{\mathrm{Q}}$s are higher in regions of projected phase--space associated with earlier infall than in regions associated with more recent infall. In particular, zones 1--3, which lie predominantly at low $R_{\rm proj}/R_{\rm vir}$ and/or low $|V_{\rm LOS}|/\sigma$ and correspond to older typical accretion times in the \citet{Pasquali2019} calibration (Figure~\ref{fig:Phase_Space}), exhibit elevated $f_{\mathrm{Q}}$s (50\%, 29\%, and 52\%, respectively). In contrast, zones 4--8, which extend to larger projected radii and/or higher line-of-sight velocities and are associated with more recent or first infall in the same mapping, show lower quenched fractions overall (23\%, 17\%, 25\%, 17\%, and 17\%, respectively). While there is scatter between adjacent zones, the lowest $f_{\mathrm{Q}}$s occur in the most recent-infall regions (zones 6--8), consistent with the visual impression in Figure~\ref{fig:Phase_Space} that star-forming systems are more prevalent at larger $R_{\rm proj}/R_{\rm vir}$ and higher $|V_{\rm LOS}|/\sigma$.

The per-zone number counts annotated in Figure~\ref{fig:QF_Phase_Space} clarify the statistical weight behind these measurements and help interpret the varying size of the uncertainty intervals. The earlier-infall zones (1--3) contain moderate sample sizes (11, 19, and 26 galaxies, respectively), whereas several of the outer/high-velocity zones contain substantially fewer galaxies (e.g. 7 and 11 galaxies in zones 7 and 8). This limited sampling contributes to broader confidence intervals and increases sensitivity to individual objects, particularly in the highest-zone indices where the PPS selection corresponds to extreme projected radii and/or velocities. Accordingly, differences between some neighbouring recent-infall zones should be interpreted cautiously, while the contrast between the earlier-infall group (zones 1--3) and the more recent-infall regime (zones 6--8) is more robust at the level supported by the current counts.

However, the bottom panel of Figure~\ref{fig:QF_Phase_Space} shows that the apparent relationship between $f_{\mathrm{Q}}$ and PPS zone is strongly dependent on stellar mass. When the sample is split into low- and high-mass galaxies, $f_{\mathrm{Q}}$ remains approximately constant for low-mass systems as they progress from early- to late-phase infall, suggesting that these galaxies may be largely unaffected by the environmental processes associated with cluster accretion, or that quenching occurs on timescales longer than their infall history. In contrast, high-mass galaxies exhibit a clear increase in $f_{\mathrm{Q}}$ from early- to late-phase cluster infall regions, indicating that their star formation is more sensitive to environmental effects such as ram-pressure stripping or tidal interactions. This divergence between low- and high-mass systems highlights the importance of stellar mass in regulating galaxy quenching and suggests that environmental quenching is most effective for massive galaxies that are already predisposed to earlier shutdown of star formation.

These PPS-based results are primarily descriptive. The mapping from PPS position to a mean infall time, $\overline{T}_{\rm inf}$, is probabilistic and subject to projection effects, redshift-space distortions, and orbital diversity, and A4038 may not perfectly match the dynamical assumptions implicit in the simulation-based calibration. In addition, the top-panel $f_{\mathrm{Q}}$ measurements do not control for stellar mass, while the bottom-panel split uses only a coarse two-bin stellar-mass partition with limited counts in some bins. The available sample sizes in the outer/high-velocity zones limit the precision of zone-to-zone comparisons. Within these limitations, the PPS analysis provides an orbit-informed, cluster-specific view showing that the $f_{\mathrm{Q}}$ varies across A4038’s projected phase--space in a manner consistent with the stratification of typical accretion history shown in Figure~\ref{fig:Phase_Space}.

\section{Discussion}
\label{sec:Discussion}

\subsection{Environmental Quenching from the SGP to the Nexus and A4038}
\label{subsec:Disc_Q}

The central aim of this work is to quantify how environment correlates with galaxy quenching, and to test whether those effects change between an ``average'' cosmological volume and a highly dynamical, assembling supercluster. Using the SGP as a reference baseline and the Nexus as a forming superstructure anchored by A4038, we find that the balance between star-forming and quenched systems depends jointly on stellar mass, halo mass.

The global $f_{\mathrm{Q}}$ measurements show clear differences between field and group environments at fixed stellar mass (Section~\ref{subsec:GQ}; Figure~\ref{fig:Binned_QF}). This separation is qualitatively consistent with satellite-specific environmental pathways that are largely absent for isolated field galaxies: removal or heating of circumgalactic gas reservoirs (``strangulation''/starvation), stripping of extended cold gas (ram pressure), and tidal interactions/harassment, all of which become more effective in deeper potentials and denser intra-group/cluster media \citep[e.g.][]{Larson1980, Gunn1972, Moore1996, Balogh2000, vanDenBosch2008, Boselli2006, Boselli2014, Peng2012, Wetzel2013, Fillingham2015}. In contrast, the strong rise of $f_\mathrm{Q}$ with stellar mass that is present across environments reflects the well-established importance of mass-linked quenching channels (e.g. structural evolution, AGN feedback, and the establishment of stable hot halos) that correlate with galaxy and halo growth \citep[e.g.][]{Kauffmann2003, Kauffmann2004, Baldry2006, Croton2006, Dekel2006, Cattaneo2006, Woo2013, Bluck2014, Bluck2016, Bluck2020}.

The inclusion of the \citet{Davies2019} comparison in Figure~\ref{fig:Binned_QF} provides critical context for understanding how systematic differences in methodology, and in particular in halo mass estimation, can substantially impact trends and resulting analysis. The halo mass estimates utilised in \citet{Davies2019} are dynamically-derived from the GAMA survey group catalogue \citep{Robotham2011}, which at low group multiplicity suffers from systematic uncertainties that typically underestimate halo masses by more than 0.5~dex at $z < 0.1$ \citep{Robotham2011, VanKempen2026}. This systematic offset is directly reflected in the comparison: \citet{Davies2019} exhibits notably higher $f_{\mathrm{Q}}$ values, especially at typical group scale halo masses ($12 < \log M_{\mathrm{halo}} < 14$), relative to our results. This is primarily due to their underestimation of the halo mass estimates, which artificially increases the $f_{\mathrm{Q}}$ at lower halo mass ranges. The halo mass improvements implemented in \citet{VanKempen2026}, upon which our work is based, directly address this concern by providing refined mass calibrations that account for these systematic biases across the full group halo mass range, with particular focus on improving accuracy within low multiplicity groups. As a result, our recovered $f_{\mathrm{Q}}$ values at low halo masses ($\log M_{\mathrm{halo}} < 14$) are systematically lower than those in \citet{Davies2019}, reflecting the true lower masses of these systems.

Importantly, at higher halo masses ($\log M_{\mathrm{halo}} \gtrsim 14$), where systems are typically more massive and group multiplicity is higher, the dynamical mass estimates used in \citet{Davies2019} become more reliable and the systematic offsets diminish. In this regime, the \citet{Davies2019} results and our measurements converge closely, indicating consistency between the two independent approaches where the dynamical mass method is most accurate. In contrast, the stellar mass trends show much closer overall agreement between our results and \citet{Davies2019}. Our dataset probes a lower stellar mass range, precluding a direct comparison of the low-mass bump, but both studies recover a steep high-mass rise in $f_{\mathrm{Q}}$ with comparable transition masses and steepness, suggesting that the stellar-mass dependence of quenching is more robust to the systematic differences in halo mass calibration that drive the offset in the halo mass panels.

A necessary step in interpreting these one-dimensional trends is to disentangle the coupled dependencies of $f_{\mathrm{Q}}$ on stellar mass and halo mass, as examining either in isolation risks misattributing trends driven by one variable to the other, given that stellar mass and halo mass are themselves correlated through the stellar-to-halo mass relation \citep[e.g.][]{Moster2010, Behroozi2013}. Figure~\ref{fig:Dissected_QF} addresses this directly. Examining $f_{\mathrm{Q}}$ as a function of halo mass at fixed stellar mass (top panels), the halo mass dependence is weakest at low stellar masses and becomes increasingly pronounced at and above $\log M_{\mathrm{stellar}} \sim 9.75$. The complementary view, $f_{\mathrm{Q}}$ as a function of stellar mass at fixed halo mass (bottom panels), reveals a strong and coherent rise with stellar mass across all halo mass bins above this threshold. These trends indicate that stellar mass sets the primary scale for quenching, while halo mass plays an increasingly important secondary role once galaxies exceed this mass range. Even the field population exhibits a rapid rise in $f_\mathrm{Q}$ across a similar stellar mass interval (Figure~\ref{fig:Dissected_QF}, bottom panel), which implies that internal or ``mass-quenching'' channels contribute strongly in this regime, plausibly linked to bulge growth, stabilisation of gas against collapse (morphological quenching), and feedback from supernovae and AGN that can suppress or prevent sustained star formation \citep[e.g.][]{Kauffmann2003, Kauffmann2004, Cattaneo2006, Croton2006, Dekel2006, Martig2009, Peng2010, Woo2013, Bluck2014, Bluck2016, Terrazas2016}.

The sharp stellar mass transition therefore marks a regime in which internal quenching becomes common, while environmental mechanisms tied to halo mass begin to exert a stronger incremental effect, steepening the observed transition in groups relative to the field. Above this mass scale, Figure~\ref{fig:Dissected_QF} shows that quenching becomes progressively more sensitive to halo mass, with higher halo mass bins exhibiting systematically elevated $f_\mathrm{Q}$ at fixed stellar mass, consistent with increasingly efficient environmental processing in deeper potentials and denser intra-halo media \citep[e.g.][]{Gunn1972, Balogh2000, vanDenBosch2008, Peng2012, Wetzel2013, Boselli2014, Fillingham2015}. A complementary, and not mutually exclusive, interpretation is that increasing halo mass correlates with the presence of stable hot gaseous haloes and longer satellite residence times, both of which raise the likelihood that satellites cross the quenching threshold once their internal regulation is already close to shutting down star formation \citep[e.g.][]{Dekel2006, Birnboim2007, Wetzel2013, Peng2015}. Together, these dissected trends suggest that the sharp stellar mass transition reflects a convergence of internal and environmental quenching channels, with halo mass modulating the efficiency of the latter above $\log M_{\mathrm{stellar}} \sim 9.75$.

A direct comparison between the SGP and Nexus in Figure~\ref{fig:Dissected_QF}, supported by the tabulated values in Tables~\ref{tab:halo_stellar_fq} and \ref{tab:stellar_halo_fq}, shows that the two environments exhibit broadly similar trends, but with measurable bin-to-bin differences that are typically modest relative to the associated uncertainties. At fixed stellar mass, the halo mass dependence of $f_{\mathrm{Q}}$ follows a consistent qualitative form in both regions, with increasing quenched fractions towards higher halo mass above $\log M_{\mathrm{stellar}} \sim 9.75$. However, the normalisation differs at the level of $\Delta f_{\mathrm{Q}} \equiv f_{\mathrm{Q}}^{\mathrm{Nexus}} - f_{\mathrm{Q}}^{\mathrm{SGP}} \sim -0.15$ to $+0.3$ across the sampled bins. The largest positive offsets (i.e. the Nexus has a larger $f_{\mathrm{Q}}$ than the SGP) are generally found at intermediate stellar masses ($9.5 \lesssim \log M_{\mathrm{stellar}} \lesssim 10.5$) and intermediate halo masses ($\log M_{\mathrm{halo}} \gtrsim 12.5$), where the Nexus can exhibit enhanced quenched fractions by $\sim 0.1$–$0.3$. In contrast, lower stellar mass bins more frequently show small negative or near-zero offsets ($\Delta f_{\mathrm{Q}} \sim -0.1$ to $0.0$), indicating comparable or slightly reduced quenching in the Nexus relative to the SGP. Importantly, these offsets are typically comparable to, or smaller than, the $1\sigma$ uncertainties reported in Tables~\ref{tab:halo_stellar_fq}, particularly in the Nexus where sample sizes are limited, and in several bins the differences are consistent with zero within the error budget.

A complementary comparison at fixed halo mass (lower panels of Figure~\ref{fig:Dissected_QF}; Table~\ref{tab:stellar_halo_fq}) reveals a similarly mixed picture. Both the SGP and Nexus show a strong and coherent increase in $f_{\mathrm{Q}}$ with stellar mass, reinforcing the dominant role of stellar mass in setting quenching probability. The differences between the two environments are again modest, typically $\Delta f_{\mathrm{Q}} \sim -0.15$ to $+0.25$, with no single halo mass bin exhibiting a systematic offset across all stellar masses. Some halo mass ranges (e.g. $\log M_{\mathrm{halo}} \sim 12.5$–$13.5$) show positive offsets at intermediate-to-high stellar masses, while others display fluctuations about zero. As within the halo-mass-sliced view, these variations are generally within the quoted uncertainties, and the scatter between bins is comparable to the measurement errors themselves.

Taken together, these results indicate that while there are hints of enhanced quenching in the Nexus in specific regions of parameter space—particularly for intermediate stellar masses in intermediate-to-high mass halos, there is no statistically robust, global offset between the SGP and Nexus at fixed stellar and halo mass. The observed differences are instead best characterised as localised variations superimposed on an otherwise consistent underlying trend. This is consistent with expectations for a dynamically assembling superstructure, in which halo mass alone may not uniquely trace environmental history due to a broader distribution of accretion times, pre-processing, and substructure \citep[e.g.][]{McGee2009, DeLucia2012, Darvish2016, Bianconi2018}. Such effects can introduce additional scatter in $f_{\mathrm{Q}}$ at fixed halo mass, particularly when averaged over limited samples.

Given the typical magnitude of the offsets ($\lesssim 0.3$) and their comparability to the uncertainties in many bins, the current dataset does not support a definitive conclusion that the Nexus systematically enhances or suppresses quenching relative to the SGP. A larger statistical sample of similar superstructures, with improved number statistics across both stellar and halo mass bins, will be required to determine whether the observed variations represent genuine environmental differences or are driven by sampling variance. For this reason, we provide the full tabulated measurements in Section~\ref{sec:Tabs} to enable direct comparison with future observational and simulation-based studies.

The A4038 PPS analysis (Figure~\ref{fig:QF_Phase_Space}) reinforces a similar point to those made prior: that stellar mass must be accounted for when interpreting environmental trends, including phase-space trends. When the PPS is analysed in coarse infall phases and split by stellar mass, the dominant contrast is between low- and high-mass galaxies: low-mass systems remain largely star-forming across Late-, Mid-, and Early-Phase bins, whereas high-mass systems are quenched at substantially higher levels in all three bins. This indicates that the PPS signal in the current sample is primarily mass-structured. The apparent enhancement in the Late-Phase high-mass bin should therefore be interpreted cautiously, as that bin contains the most massive galaxies and the stellar-mass split is necessarily coarse. With finer high-mass binning and larger counts, part of that residual phase dependence may weaken. At present, we therefore treat the PPS result as consistent with stellar-mass-driven quenching modulated by environment, rather than as definitive evidence for an independent phase effect at fixed stellar mass.

These results build directly on \citetalias{VanKempen2024}, which measured SGP $f_{\mathrm{Q}}$ trends using group membership as the environmental metric and found a strong environmental dependence at fixed stellar mass. The present analysis largely recovers the same qualitative behaviour, while demonstrating that replacing membership-based binning with group-scale halo masses from \citet{VanKempen2026} reduces the blending of physically distinct environments. In particular, membership bins can conflate a non-negligible range of halo masses (especially at low group memberships), which can smooth or dilute underlying trends; binning by halo mass instead provides a more directly comparable measure of potential depth and intra-halo conditions, and therefore a cleaner statistical separation of quenching behaviour across environments.

\subsection{Star Formation: regulation and variability}
\label{subsec:Disc_SF}

This work also constrains how environment regulates star formation within the star-forming population, by comparing the mean and standard deviation of $\log \mathrm{sSFR}$ across the SGP baseline and the dynamically active Nexus region. Globally, the mean $\log \mathrm{sSFR}$ declines with increasing stellar mass in both field and group environments (Section~\ref{subsec:GSF}; Figure~\ref{fig:Binned_Mean}), consistent with the established mass dependence of the star-forming main sequence \citep[e.g.][]{Noeske2007, Whitaker2012, Speagle2014}. For group galaxies, an additional decline of mean $\log \mathrm{sSFR}$ with increasing halo mass is present in the aggregate relations in both SGP and Nexus (Figure~\ref{fig:Binned_Mean}), indicating that environment can suppress star formation even among galaxies that remain classified as star-forming.

The de-coupled trends in Figure~\ref{fig:Dissected_Mean} are central to interpreting this behaviour, as they separate halo mass and stellar mass effects that are otherwise covariant in one-dimensional projections. At fixed stellar mass, the mean $\log \mathrm{sSFR}$ generally declines towards higher halo mass, while at fixed halo mass it decreases strongly with stellar mass. This confirms that stellar mass remains the primary determinant of typical star-forming activity, with halo mass introducing a secondary, environment-linked suppression. In physical terms, such a halo mass dependence is consistent with processes that reduce gas supply or star formation efficiency without immediately quenching galaxies, including reduced cosmological accretion, stripping or heating of circumgalactic gas (``strangulation''), and interaction with a denser intra-halo medium in more massive systems \citep[e.g.][]{Larson1980, Balogh2000, vanDenBosch2008, Boselli2014, Peng2015}. The strength of this suppression varies across stellar mass bins, indicating a mass-dependent susceptibility to environmental regulation rather than a uniform effect.

The observed scatter of this distribution $(\sigma\log \mathrm{sSFR})$, provides further insight. At low stellar masses ($\log M_\mathrm{stellar}<10$), the scatter is modest ($\sim$0.1--0.4 dex), while at higher stellar masses it increases substantially (often $\gtrsim$0.6 dex; Figure~\ref{fig:Global_SF}). This indicates that the diversity of star-forming states is primarily linked to stellar mass rather than halo mass. This is consistent with the idea that ``bursty'' star formation in low-mass galaxies produces short-timescale variability that is partially averaged out by SFR tracers and sample selection \citep[e.g.][]{McQuinn2010, Weisz2012, Kennicutt2012}, while at higher stellar masses the star-forming population becomes increasingly heterogeneous as galaxies approach quenching, amplifying both intrinsic variation and observational uncertainties \citep[e.g.][]{Whitaker2012, Speagle2014, Tacconi2020}.

The two-dimensional view in Figure~\ref{fig:Global_SF} reinforces that $\log \mathrm{sSFR}$ varies across the joint stellar mass--halo mass plane, and that marginalised relations can conflate these coupled dependencies. Notably, $\sigma\log \mathrm{sSFR}$ varies much more strongly with stellar mass than with halo mass, suggesting that environment primarily shifts the typical level of star formation rather than strongly broadening the distribution of star-forming activity. This is consistent with models in which internal gas-cycle processes set the range of star formation states, while environmental effects act more as a systematic suppression of the mean \citep[e.g.][]{Saintonge2017, Catinella2018, Tacconi2020}.

A direct comparison of the mean $\log \mathrm{sSFR}$ between the SGP and Nexus, using the de-coupled trends in Figure~\ref{fig:Dissected_Mean} together with Tables~\ref{tab:halo_stellar_sSFR} and \ref{tab:stellar_halo_sSFR}, shows that the two environments follow closely similar trends, but with measurable, bin-dependent offsets. At fixed stellar mass, the halo mass dependence is broadly consistent between the SGP and Nexus, with both showing declining $\log \mathrm{sSFR}$ towards higher halo mass. The differences are typically modest, with $\Delta \log \mathrm{sSFR} \sim -0.15$ to $+0.1$ dex. Negative offsets (i.e. indicating lower $\log \mathrm{sSFR}$ in the Nexus) are most commonly found at intermediate stellar masses ($9 \lesssim \log M_{\mathrm{stellar}} \lesssim 10.5$) and intermediate group halo masses ($\log M_{\mathrm{halo}} \gtrsim 12.5$), where suppressions of $\sim 0.1$–$0.2$ dex are observed. At lower halo masses and lower stellar masses, the offsets are smaller and often consistent with zero within the uncertainties reported in Table~\ref{tab:halo_stellar_sSFR}.

As a function of stellar mass at fixed halo mass (lower panels of Figure~\ref{fig:Dissected_Mean}; Table~\ref{tab:stellar_halo_sSFR}) shows a similarly consistent stellar-mass dependence in both environments, with $\log \mathrm{sSFR}$ declining strongly with increasing stellar mass. The differences between the SGP and Nexus again remain modest, typically $\Delta \log \mathrm{sSFR} \sim -0.2$ to $+0.1$ dex, with the largest negative offsets appearing at intermediate-to-high stellar masses within moderately massive haloes ($\log M_{\mathrm{halo}} \sim 12.5$–$13.5$). However, these offsets are not monotonic across all halo mass bins, and several bins show fluctuations about zero or reversals in sign. As in the halo-mass-sliced view, the magnitude of the differences is comparable to the measurement uncertainties, indicating that many bins are consistent within errors.

Taken together, these results indicate that the SGP and Nexus share a common underlying dependence of sSFR with stellar and halo mass, with only secondary, localised differences in normalisation. There are indications that star formation may be modestly more suppressed in the Nexus in certain regions of parameter space, particularly for intermediate stellar mass galaxies in intermediate to large mass halos, but these offsets are typically $\lesssim 0.2$ dex and are not consistently present across all bins. This behaviour is consistent with expectations for a dynamically assembling superstructure, in which a broader distribution of accretion histories and pre-processing pathways can introduce additional scatter at fixed halo mass \citep[e.g.][]{McGee2009, Wetzel2013, Haines2015, Darvish2016}.

As with the $f_{\mathrm{Q}}$ results, given the relatively small magnitude of the offsets in sSFRand their comparability to the uncertainties, the current dataset does not support a definitive conclusion that the Nexus systematically suppresses or enhances star formation relative to the SGP at fixed stellar and halo mass. Instead, the results are best interpreted as broadly consistent behaviour with indications of increased variability in the Nexus. A larger statistical sample, particularly of Nexus-like structures, will be required to determine whether these differences reflect genuine environmental modulation or arise from sampling variance. For this reason, we provide the full tabulated measurements in Section~\ref{sec:Tabs} to enable direct comparison with future studies.

The three radial zone distributions in Figure~\ref{fig:ssfr_zones} show that at different distances to the large-scale over-density, they largely probe different parts of the stellar mass--halo mass plane, particularly in halo mass coverage. Consequently, the apparent zone-to-zone differences are most naturally explained by variations in the halo mass function sampled in each zone, together with small-number statistics, rather than by projected distance from A4038 alone. With the present dataset, we therefore cannot claim a robust distance-driven modulation at fixed stellar and halo mass.

Several limitations shape the strength of these inferences. The Nexus sample is much smaller than the SGP, making the zonal results primarily qualitative, particularly in sparsely populated regions of parameter space, and $\sigma~\log~\mathrm{sSFR}$ necessarily combines intrinsic diversity with measurement uncertainty. In addition, KDE smoothing and masking limit sensitivity to fine-scale features, and redshift-space distortions (e.g. the Kaiser effect) complicate precise distance-based environmental classifications. 

These results also connect to \citetalias{VanKempen2024}, which reported a tentative deficit in star formation for star-forming group galaxies relative to the field in the SGP when using group membership as the environment metric, but with substantial scatter. With the improved halo masses introduced in \citet{VanKempen2026}, separating by halo mass reduces environmental mixing and reveals a clearer separation between environments, while retaining the dominant stellar mass dependence of $\log \mathrm{sSFR}$. 

Progress towards a causal interpretation will require both improved baryonic tracers and more precise environmental characterisation. Measurements of H\,\textsc{I} and molecular gas are particularly important for distinguishing reduced gas supply from direct gas removal, and for linking suppression in $\log \mathrm{sSFR}$ to gas fractions and depletion times \citep[e.g.][]{Boselli2014, Saintonge2017, Catinella2018, Tacconi2020}. Improved halo mass estimates \citep[e.g.][]{VanKempen2026} and more physically motivated environmental metrics will further reduce systematic uncertainties.

\section{Summary \& Conclusions}
\label{sec:Sum_Con}

Environment is a fundamental driver of galaxy evolution in the local Universe, contributing to how, when and where galaxies cease forming stars and how rapidly star formation is regulated as structures assemble hierarchically from the field, into groups, filaments, and clusters. By combining an ``average'' reference volume (SGP) with a dynamically active forming superstructure (the Nexus, centred on A4038), this study demonstrates that environmental influence is not captured by a single scalar metric, but depends jointly on stellar mass, group-scale halo mass, and a galaxy’s location within the surrounding large-scale structure. Placing A4038 in this broader context is therefore essential: the cluster’s galaxy population reflects not only in-situ cluster physics, but also the accretion and pre-processing histories encoded by the larger environment in which the cluster is embedded.

We summarise our main findings as:

\begin{itemize}
\item Using a large ``global'' SGP sample as a reference baseline and compared it to the dynamically evolving Nexus region surrounding A4038, enabling an explicit test of how galaxy evolution differs between average and supercluster-like environments.

\item Quantified environmental quenching using the $f_\mathrm{Q}$ (quenched fraction), measured as a function of stellar mass and (for grouped galaxies) halo mass. We summarised one-dimensional trends using logistic fits parameterised by $M_{50}$ and $k$, and we used controlled/dissected measurements to separate halo mass and stellar mass effects.

\item Globally, $f_\mathrm{Q}$ increases with stellar mass in both group and field environments, and (for group galaxies) tends to increase with halo mass. The dissected analysis shows that marginalised relations can conflate coupled dependencies, and that controlled trends are necessary to interpret how halo mass and stellar mass jointly regulate quenching.

\item The Nexus group population shows quenching behaviour that shows marginal variations from the SGP baseline and generally consistent when considering uncertainties, is consistent with the expectation that a forming superstructure mixes a broader range of accretion histories and pre-processing pathways at fixed present-day halo mass.

\item Characterised star formation within the star-forming population by measuring the mean and standard deviation of $\log \mathrm{sSFR}$ across stellar mass and halo mass; mean $\log \mathrm{sSFR}$ declines with increasing stellar mass across environments, with additional environment-linked suppression in group-scale halos that is clearest when controlling for stellar mass. The diversity of star-forming activity (the scatter) varies most strongly with stellar mass and shows comparatively weaker dependence on halo mass over the range probed.

\item Within the Nexus, we examined radial trends around A4038 and the forming supercluster using three radial zones. Both $f_\mathrm{Q}$ and star-forming activity demonstrated zone-to-zone variation, largely due to differing sampling of the halo mass function demonstrating that the regulatory processes of star formation may vary across the forming superstructure.

\item For A4038, we used a projected phase-space analysis to connect quenching to accretion history. Quenched fractions are higher in projected phase-space regions associated with earlier infall. However, these are shown to be heavily stellar mass dependent with low mass galaxies not exhibiting the same characteristics.

\item The SGP and Nexus show consistent underlying dependencies of $f_{\mathrm{Q}}$ and $\log \mathrm{sSFR}$ on stellar and halo mass, with only modest offsets ($\Delta f_{\mathrm{Q}} \lesssim 0.3$, $\Delta \log \mathrm{sSFR} \lesssim 0.2$ dex) across the parameter space. These differences are generally comparable to the uncertainties, and we find no definitive evidence that the Nexus systematically enhances or suppresses quenching or star formation at fixed stellar and halo mass.

\item Key limitations include restricted sample sizes (particularly within Nexus sub-regions and extreme projected phase--space regimes), projection and redshift--space distortions, and the finite resolution imposed by KDE smoothing and masking. These factors motivate the use of larger, more homogeneous spectroscopic samples and the incorporation of dynamical and environmental kinematic information in future analyses.
\end{itemize}

Looking ahead, forthcoming wide-area spectroscopic surveys in the southern hemisphere will enable a step change in this kind of analysis. Highly complete redshift surveys such as 4HS \citep{ENTaylor2023} and WAVES \citep{Driver2019} will provide the statistical power and uniform environmental metrics required to extend controlled measurements (at fixed stellar mass and halo mass) across all types of galaxy environments. When combined with wide-area multi-wavelength datasets tracing the baryon cycle, including X-ray mapping of the hot ICM and CGM with eROSITA, H\,\textsc{I} surveys with ASKAP and SKA, mid-IR measurements from WISE that trace dust-obscured star formation and stellar mass, deep optical spectroscopy from facilities such as the VLT to constrain ionised-gas and AGN diagnostics and stellar populations, optical imaging from LSST for galaxy structure and weak-lensing mass mapping, and near-IR imaging from VHS and Euclid for robust stellar masses and dust-reduced morphologies, the community will be able to build an unprecedented panchromatic view of how gas supply, gas removal, feedback, and structure growth connect environment to star formation and quenching. In this sense, the present work provides both (i) an empirical foundation showing where environmental effects are most apparent in stellar- and halo mass space, and (ii) a roadmap for where targeted multi-wavelength follow-up can most directly test quenching and star-formation regulation mechanisms.

\section{Acknowledgements}
\label{sec:Acknowledgements}

We thank the anonymous referee for helpful comments and suggestions that have improved the content and clarity of this paper. M.E.C. is a recipient of an Australian Research Council Future Fellowship (project No. FT170100273) funded by the Australian Government. D.J.C. is a recipient of an Australian Research Council Future Fellowship (project No. FT220100841) funded by the Australian Government. This publication makes use of data products from the Wide-field Infrared Survey Explorer, which is a joint project of the University of California, Los Angeles, and the Jet Propulsion Laboratory/California Institute of Technology, funded by the National Aeronautics and Space Administration. GAMA is a joint European-Australasian project based around a spectroscopic campaign using the Anglo-Australian Telescope. The GAMA input catalogue is based on data taken from the Sloan Digital Sky Survey and the UKIRT Infrared Deep Sky Survey. Complementary imaging of the GAMA regions is being obtained by a number of independent survey programmes including GALEX MIS, VST KiDS, VISTA VIKING, WISE, Herschel-ATLAS, GMRT and ASKAP providing UV to radio coverage. GAMA is funded by the STFC (UK), the ARC (Australia), the AAO, and the participating institutions. The GAMA website is \url{https://www.gama-survey.org/}. Based on observations made with ESO Telescopes at the La Silla Paranal Observatory under programme ID 177.A-3016. This research has made use of {\tt python} (\url{https://www.python.org}) and python packages: {\tt astropy} \citep{Astropy2013,Astropy2018, Astropy2022}, {\tt cmasher} \citep{cmasher}, {\tt matplotlib} \url{http://matplotlib.org/} \citep{Hunter2007}, {\tt NumPy} \url{http://www.numpy.org/} \citep{Walt2011}, {\tt Pandas} \citep{Pandas}, and {\tt SciPy} \url{https://www.scipy.org/} \citep{Virtanen2020}.

\bibliography{ref}

\appendix
\section{Table Statistics}
\label{sec:Tabs}

\renewcommand{\thetable}{A\arabic{table}}
\renewcommand{\thefigure}{A\arabic{figure}}


\begin{table*}
\centering
\caption{Quenched fraction $f_{\mathrm{Q}}$ for 0.4 dex bins of halo mass (top) and 0.2 dex bins of stellar mass (bottom) for the SGP and Nexus, for both group and field galaxies. Values are the bootstrapped median with uncertainties spanning the 16th and 84th percentiles. SGP bins with fewer than ten galaxies and Nexus bins fewer than five galaxies are indicated by ``--'' and errors are 16$^{th}$ and 84$^{th}$ percentiles from the bootstrapping.}
\label{tab:fq_binned}
\resizebox{\textwidth}{!}{%
\begin{tabular}{ccccc}
\hline
$M_{\mathrm{halo}}~[\log M_{\odot}]$ & $f_{\mathrm{Q}}$ SGP (Group) & $f_{\mathrm{Q}}$ Nexus (Group) & $f_{\mathrm{Q}}$ SGP (Field) & $f_{\mathrm{Q}}$ Nexus (Field) \\
\hline
11.6 & $0.21^{+0.06}_{-0.06}$ & $0.18^{+0.09}_{-0.09}$ & -- & -- \\
12.0 & $0.09^{+0.02}_{-0.02}$ & $0.04^{+0.04}_{-0.04}$ & -- & -- \\
12.4 & $0.16^{+0.01}_{-0.01}$ & $0.14^{+0.05}_{-0.05}$ & -- & -- \\
12.8 & $0.24^{+0.01}_{-0.01}$ & $0.29^{+0.05}_{-0.05}$ & -- & -- \\
13.2 & $0.31^{+0.01}_{-0.01}$ & $0.20^{+0.03}_{-0.03}$ & -- & -- \\
13.6 & $0.35^{+0.01}_{-0.01}$ & $0.20^{+0.03}_{-0.03}$ & -- & -- \\
14.0 & $0.41^{+0.02}_{-0.02}$ & $0.28^{+0.03}_{-0.03}$ & -- & -- \\
14.4 & $0.67^{+0.04}_{-0.04}$ & -- & -- & -- \\
\hline
$M_{\mathrm{stellar}}~[\log M_{\odot}]$ & $f_{\mathrm{Q}}$ SGP (Group) & $f_{\mathrm{Q}}$ Nexus (Group) & $f_{\mathrm{Q}}$ SGP (Field) & $f_{\mathrm{Q}}$ Nexus (Field) \\
\hline
8.2  & $0.52^{+0.09}_{-0.09}$ & $0.00^{+0.00}_{-0.00}$ & $0.28^{+0.07}_{-0.07}$ & $0.00^{+0.00}_{-0.00}$ \\
8.4  & $0.17^{+0.06}_{-0.06}$ & $0.02^{+0.02}_{-0.02}$ & $0.30^{+0.05}_{-0.05}$ & $0.22^{+0.06}_{-0.06}$ \\
8.6  & $0.20^{+0.04}_{-0.04}$ & $0.14^{+0.06}_{-0.06}$ & $0.26^{+0.03}_{-0.03}$ & $0.29^{+0.06}_{-0.06}$ \\
8.8  & $0.14^{+0.03}_{-0.03}$ & $0.09^{+0.04}_{-0.04}$ & $0.18^{+0.02}_{-0.02}$ & $0.17^{+0.06}_{-0.06}$ \\
9.0  & $0.14^{+0.02}_{-0.02}$ & $0.08^{+0.03}_{-0.03}$ & $0.21^{+0.02}_{-0.02}$ & $0.24^{+0.08}_{-0.08}$ \\
9.2  & $0.19^{+0.02}_{-0.02}$ & $0.04^{+0.02}_{-0.02}$ & $0.16^{+0.01}_{-0.01}$ & $0.06^{+0.04}_{-0.04}$ \\
9.4  & $0.17^{+0.01}_{-0.01}$ & $0.11^{+0.04}_{-0.04}$ & $0.10^{+0.01}_{-0.01}$ & $0.00^{+0.00}_{-0.00}$ \\
9.6  & $0.16^{+0.01}_{-0.01}$ & $0.00^{+0.00}_{-0.00}$ & $0.09^{+0.01}_{-0.01}$ & $0.00^{+0.00}_{-0.00}$ \\
9.8  & $0.09^{+0.01}_{-0.01}$ & $0.02^{+0.02}_{-0.02}$ & $0.07^{+0.01}_{-0.01}$ & $0.00^{+0.00}_{-0.00}$ \\
10.0 & $0.18^{+0.01}_{-0.01}$ & $0.37^{+0.06}_{-0.06}$ & $0.08^{+0.01}_{-0.01}$ & $0.19^{+0.07}_{-0.07}$ \\
10.2 & $0.35^{+0.02}_{-0.02}$ & $0.50^{+0.08}_{-0.08}$ & $0.20^{+0.01}_{-0.01}$ & $0.41^{+0.11}_{-0.11}$ \\
10.4 & $0.42^{+0.02}_{-0.02}$ & $0.66^{+0.09}_{-0.09}$ & $0.28^{+0.01}_{-0.01}$ & $0.17^{+0.08}_{-0.08}$ \\
10.6 & $0.48^{+0.02}_{-0.02}$ & $0.58^{+0.10}_{-0.10}$ & $0.38^{+0.02}_{-0.02}$ & $0.54^{+0.15}_{-0.15}$ \\
10.8 & $0.61^{+0.02}_{-0.02}$ & $0.76^{+0.08}_{-0.08}$ & $0.43^{+0.02}_{-0.02}$ & -- \\
11.0 & $0.71^{+0.03}_{-0.03}$ & $0.85^{+0.08}_{-0.08}$ & $0.53^{+0.03}_{-0.04}$ & -- \\
11.2 & $0.85^{+0.03}_{-0.03}$ & -- & $0.69^{+0.06}_{-0.06}$ & -- \\
\hline
\end{tabular}
}
\end{table*}

\begin{table*}
\centering
\caption{Quenched fraction ($f_Q$) of galaxies as a function of halo mass for both the SGP and Nexus group samples. $f_Q$ values are the bootstrapped median values. Bins with fewer than three galaxies (or missing data) are indicated by ``--'' and errors are 16$^{th}$ and 84$^{th}$ percentiles from the bootstrapping.}
\label{tab:halo_stellar_fq}
\resizebox{\textwidth}{!}{%
\begin{tabular}{c|cccccccccccc}
\hline
$M_{\mathrm{stellar}}~[\log M_{\odot}]$ & \multicolumn{12}{c}{$M_{\mathrm{halo}}$ [$\log M_{\odot}$]} \\
\hline
 & \multicolumn{2}{c}{11.75} & \multicolumn{2}{c}{12.25} & \multicolumn{2}{c}{12.75} & \multicolumn{2}{c}{13.25} & \multicolumn{2}{c}{13.75} & \multicolumn{2}{c}{14.25} \\
 & SGP & Nexus & SGP & Nexus & SGP & Nexus & SGP & Nexus & SGP & Nexus & SGP & Nexus \\
\hline
8.5--9   & 
$0.10^{+0.04}_{-0.05}$ & $0.22^{+0.18}_{-0.12}$ & 
$0.23^{+0.04}_{-0.07}$ & $0.29^{+0.16}_{-0.13}$ & 
$0.20^{+0.04}_{-0.04}$ & $0.15^{+0.07}_{-0.07}$ & 
$0.15^{+0.03}_{-0.03}$ & $0.13^{+0.05}_{-0.07}$ & 
$0.06^{+0.03}_{-0.02}$ & $0.02^{+0.02}_{-0.02}$ & 
$0.60^{+0.20}_{-0.27}$ & -- \\
9--9.5   & 
$0.08^{+0.04}_{-0.04}$ & $0.00^{+0.00}_{-0.00}$ & 
$0.12^{+0.03}_{-0.02}$ & $0.00^{+0.00}_{-0.00}$ & 
$0.13^{+0.02}_{-0.02}$ & $0.05^{+0.06}_{-0.05}$ & 
$0.19^{+0.02}_{-0.02}$ & $0.05^{+0.04}_{-0.03}$ & 
$0.18^{+0.02}_{-0.02}$ & $0.09^{+0.04}_{-0.03}$ & 
$0.40^{+0.07}_{-0.06}$ & -- \\
9.5--10  & 
$0.00^{+0.00}_{-0.00}$ & -- & 
$0.09^{+0.01}_{-0.02}$ & $0.06^{+0.09}_{-0.06}$ & 
$0.06^{+0.01}_{-0.01}$ & $0.00^{+0.00}_{-0.00}$ & 
$0.11^{+0.01}_{-0.01}$ & $0.02^{+0.03}_{-0.02}$ & 
$0.15^{+0.02}_{-0.01}$ & $0.08^{+0.03}_{-0.02}$ & 
$0.30^{+0.05}_{-0.04}$ & -- \\
10--10.5 & 
-- & -- & 
$0.27^{+0.05}_{-0.04}$ & $0.38^{+0.16}_{-0.15}$ & 
$0.30^{+0.02}_{-0.02}$ & $0.48^{+0.13}_{-0.08}$ & 
$0.31^{+0.02}_{-0.02}$ & $0.41^{+0.09}_{-0.12}$ & 
$0.44^{+0.02}_{-0.02}$ & $0.69^{+0.06}_{-0.07}$ & 
$0.70^{+0.04}_{-0.05}$ & -- \\
10.5--11 & 
-- & -- & 
-- & -- & 
$0.47^{+0.02}_{-0.03}$ & $0.86^{+0.07}_{-0.13}$ & 
$0.55^{+0.02}_{-0.02}$ & $0.58^{+0.11}_{-0.08}$ & 
$0.61^{+0.03}_{-0.03}$ & $0.72^{+0.08}_{-0.10}$ & 
$0.83^{+0.04}_{-0.05}$ & -- \\
11--11.5 & 
-- & -- & 
-- & -- & 
-- & -- & 
$0.74^{+0.03}_{-0.03}$ & $0.78^{+0.22}_{-0.28}$ & 
$0.85^{+0.03}_{-0.03}$ & $0.86^{+0.14}_{-0.14}$ & 
$0.88^{+0.06}_{-0.09}$ & -- \\
11.5--12 & 
-- & -- & 
-- & -- & 
-- & -- & 
-- & -- & 
$1.00^{+0.00}_{-0.00}$ & -- & 
$1.00^{+0.00}_{-0.00}$ & -- \\
\hline
\end{tabular}
}
\end{table*}

\begin{table*}
\centering
\caption{Quenched fraction ($f_Q$) of galaxies as a function of stellar mass for different halo mass bins for both the SGP and Nexus samples. $f_Q$ values are the bootstrapped median values. Bins with fewer than three galaxies (or missing data) are indicated by ``--'' and errors are 16$^{th}$ and 84$^{th}$ percentiles from the bootstrapping.}
\label{tab:stellar_halo_fq}
\resizebox{\textwidth}{!}{%
\begin{tabular}{c|cccccccccccccc}
\hline
$M_{\mathrm{halo}} \, [\log M_{\odot}]$ & \multicolumn{14}{c}{$M_\mathrm{stellar}$ [$\log M_{\odot}$]} \\
\hline
  & \multicolumn{2}{c}{8.75} & \multicolumn{2}{c}{9.25} & \multicolumn{2}{c}{9.75} & \multicolumn{2}{c}{10.25} & \multicolumn{2}{c}{10.75} & \multicolumn{2}{c}{11.25} & \multicolumn{2}{c}{11.75} \\
  & SGP & Nexus & SGP & Nexus & SGP & Nexus & SGP & Nexus & SGP & Nexus & SGP & Nexus & SGP & Nexus \\
\hline
Field      &
$0.21^{+0.01}_{-0.01}$ & $0.23^{+0.04}_{-0.04}$ &
$0.13^{+0.01}_{-0.01}$ & $0.08^{+0.03}_{-0.03}$ &
$0.07^{+0.00}_{-0.00}$ & $0.01^{+0.01}_{-0.01}$ &
$0.21^{+0.01}_{-0.01}$ & $0.31^{+0.06}_{-0.06}$ &
$0.41^{+0.01}_{-0.01}$ & $0.58^{+0.11}_{-0.11}$ &
$0.61^{+0.04}_{-0.04}$ & -- &
-- & -- \\
11.5--12   &
$0.10^{+0.04}_{-0.05}$ & $0.22^{+0.18}_{-0.12}$ &
$0.08^{+0.04}_{-0.04}$ & $0.00^{+0.00}_{-0.00}$ &
$0.00^{+0.00}_{-0.00}$ & -- &
-- & -- &
-- & -- &
-- & -- &
-- & -- \\
12--12.5   &
$0.23^{+0.04}_{-0.07}$ & $0.29^{+0.16}_{-0.13}$ &
$0.12^{+0.03}_{-0.02}$ & $0.00^{+0.00}_{-0.00}$ &
$0.09^{+0.01}_{-0.02}$ & $0.06^{+0.09}_{-0.06}$ &
$0.27^{+0.05}_{-0.04}$ & $0.38^{+0.16}_{-0.15}$ &
-- & -- &
-- & -- &
-- & -- \\
12.5--13   &
$0.20^{+0.04}_{-0.04}$ & $0.15^{+0.07}_{-0.07}$ &
$0.13^{+0.02}_{-0.02}$ & $0.05^{+0.06}_{-0.05}$ &
$0.06^{+0.01}_{-0.01}$ & $0.00^{+0.00}_{-0.00}$ &
$0.30^{+0.02}_{-0.02}$ & $0.48^{+0.13}_{-0.08}$ &
$0.47^{+0.02}_{-0.03}$ & $0.86^{+0.07}_{-0.13}$ &
-- & -- &
-- & -- \\
13--13.5   &
$0.15^{+0.03}_{-0.03}$ & $0.13^{+0.05}_{-0.07}$ &
$0.19^{+0.02}_{-0.02}$ & $0.05^{+0.04}_{-0.03}$ &
$0.11^{+0.01}_{-0.01}$ & $0.02^{+0.03}_{-0.02}$ &
$0.31^{+0.02}_{-0.02}$ & $0.41^{+0.09}_{-0.12}$ &
$0.55^{+0.02}_{-0.02}$ & $0.58^{+0.11}_{-0.08}$ &
$0.74^{+0.03}_{-0.03}$ & $0.78^{+0.22}_{-0.28}$ &
-- & -- \\
13.5--14   &
$0.06^{+0.03}_{-0.02}$ & $0.02^{+0.02}_{-0.02}$ &
$0.18^{+0.02}_{-0.02}$ & $0.09^{+0.04}_{-0.03}$ &
$0.15^{+0.02}_{-0.01}$ & $0.08^{+0.03}_{-0.02}$ &
$0.44^{+0.02}_{-0.02}$ & $0.69^{+0.06}_{-0.07}$ &
$0.61^{+0.03}_{-0.03}$ & $0.72^{+0.08}_{-0.10}$ &
$0.85^{+0.03}_{-0.03}$ & $0.86^{+0.14}_{-0.14}$ &
$1.00^{+0.00}_{-0.00}$ & -- \\
14--14.5   &
$0.60^{+0.20}_{-0.27}$ & -- &
$0.40^{+0.07}_{-0.06}$ & -- &
$0.30^{+0.05}_{-0.04}$ & -- &
$0.70^{+0.04}_{-0.05}$ & -- &
$0.83^{+0.04}_{-0.05}$ & -- &
$0.88^{+0.06}_{-0.09}$ & -- &
$1.00^{+0.00}_{-0.00}$ & -- \\
\hline
\end{tabular}
}
\end{table*}

\begin{table*}
\centering
\caption{Bootstrapped medians of the mean $\log~\mathrm{sSFR}~[\log~\mathrm{yr}^{-1}]$ for 0.4 dex bins of halo mass (top) and 0.2 dex bins of stellar mass (bottom) for the SGP and Nexus, for both group and field galaxies. Uncertainties span the 16th and 84th percentiles. SGP bins with fewer than eight galaxies and Nexus bins fewer than four galaxies are indicated by ``--'' and errors are 16$^{th}$ and 84$^{th}$ percentiles from the bootstrap resampling.}
\label{tab:ssfr_binned}
\resizebox{\textwidth}{!}{%
\begin{tabular}{ccccc}
\hline
$M_{\mathrm{halo}}~[\log M_{\odot}]$ & $\log \mathrm{sSFR}$ SGP (Group) & $\log \mathrm{sSFR}$ Nexus (Group) & $\log \mathrm{sSFR}$ SGP (Field) & $\log \mathrm{sSFR}$ Nexus (Field) \\
\hline
11.6 & $-9.74^{+0.09}_{-0.08}$ & -- & -- & -- \\
12.0 & $-9.80^{+0.03}_{-0.03}$ & $-9.94^{+0.10}_{-0.12}$ & -- & -- \\
12.4 & $-9.99^{+0.02}_{-0.02}$ & $-10.18^{+0.10}_{-0.11}$ & -- & -- \\
12.8 & $-10.14^{+0.02}_{-0.02}$ & $-10.21^{+0.09}_{-0.10}$ & -- & -- \\
13.2 & $-10.30^{+0.02}_{-0.02}$ & $-10.28^{+0.07}_{-0.07}$ & -- & -- \\
13.6 & $-10.40^{+0.02}_{-0.02}$ & $-10.30^{+0.07}_{-0.08}$ & -- & -- \\
14.0 & $-10.41^{+0.03}_{-0.03}$ & $-10.32^{+0.09}_{-0.09}$ & -- & -- \\
14.4 & $-10.89^{+0.12}_{-0.12}$ & -- & -- & -- \\
\hline
$M_{\mathrm{stellar}}~[\log M_{\odot}]$ & $\log \mathrm{sSFR}$ SGP (Group) & $\log \mathrm{sSFR}$ Nexus (Group) & $\log \mathrm{sSFR}$ SGP (Field) & $\log \mathrm{sSFR}$ Nexus (Field) \\
\hline
8.2  & -- & -- & $-9.64^{+0.05}_{-0.05}$ & -- \\
8.4  & -- & -- & $-9.69^{+0.07}_{-0.07}$ & -- \\
8.6  & $-9.62^{+0.22}_{-0.18}$ & -- & $-9.67^{+0.04}_{-0.04}$ & $-9.63^{+0.05}_{-0.05}$ \\
8.8  & $-9.71^{+0.05}_{-0.05}$ & $-9.74^{+0.08}_{-0.08}$ & $-9.75^{+0.04}_{-0.04}$ & $-9.68^{+0.06}_{-0.06}$ \\
9.0  & $-9.70^{+0.03}_{-0.03}$ & $-9.68^{+0.03}_{-0.03}$ & $-9.75^{+0.03}_{-0.03}$ & $-9.94^{+0.07}_{-0.07}$ \\
9.2  & $-9.83^{+0.02}_{-0.02}$ & $-9.87^{+0.04}_{-0.05}$ & $-9.78^{+0.02}_{-0.02}$ & $-9.85^{+0.04}_{-0.05}$ \\
9.4  & $-9.80^{+0.02}_{-0.01}$ & $-9.86^{+0.02}_{-0.02}$ & $-9.78^{+0.01}_{-0.01}$ & $-9.85^{+0.03}_{-0.04}$ \\
9.6  & $-9.84^{+0.01}_{-0.01}$ & $-9.93^{+0.03}_{-0.03}$ & $-9.85^{+0.01}_{-0.01}$ & $-9.94^{+0.02}_{-0.03}$ \\
9.8  & $-9.90^{+0.01}_{-0.01}$ & $-10.01^{+0.03}_{-0.03}$ & $-9.87^{+0.01}_{-0.01}$ & $-9.96^{+0.02}_{-0.02}$ \\
10.0 & $-10.10^{+0.01}_{-0.01}$ & $-10.45^{+0.11}_{-0.11}$ & $-10.04^{+0.01}_{-0.01}$ & $-10.38^{+0.09}_{-0.10}$ \\
10.2 & $-10.21^{+0.02}_{-0.02}$ & $-10.69^{+0.11}_{-0.12}$ & $-10.13^{+0.01}_{-0.01}$ & $-10.45^{+0.13}_{-0.15}$ \\
10.4 & $-10.43^{+0.02}_{-0.02}$ & $-10.88^{+0.15}_{-0.16}$ & $-10.30^{+0.01}_{-0.01}$ & $-10.21^{+0.10}_{-0.09}$ \\
10.6 & $-10.62^{+0.03}_{-0.03}$ & $-10.78^{+0.12}_{-0.13}$ & $-10.47^{+0.02}_{-0.02}$ & $-10.77^{+0.21}_{-0.21}$ \\
10.8 & $-10.93^{+0.03}_{-0.03}$ & $-11.10^{+0.17}_{-0.17}$ & $-10.68^{+0.03}_{-0.03}$ & -- \\
11.0 & $-11.01^{+0.05}_{-0.05}$ & $-11.17^{+0.21}_{-0.21}$ & $-10.84^{+0.05}_{-0.05}$ & -- \\
11.2 & $-11.25^{+0.07}_{-0.07}$ & -- & $-11.13^{+0.08}_{-0.09}$ & -- \\
\hline
\end{tabular}
}
\end{table*}


\begin{table*}
\centering
\caption{Mean $\log \mathrm{sSFR}~[\log \mathrm{yr}^{-1}]$ of galaxies as a function of halo mass for different stellar mass bins for both the SGP and Nexus samples. Values are the bootstrapped medians, with uncertainties spanning the 16th and 84th percentiles. Bins with fewer than three galaxies are indicated by ``--''.}
\label{tab:halo_stellar_sSFR}
\resizebox{\textwidth}{!}{%
\begin{tabular}{c|cccccccccccc}
\hline
$M_{\mathrm{stellar}}~[\log M_{\odot}]$ & \multicolumn{12}{c}{$M_{\mathrm{halo}}$ [$\log M_{\odot}$]} \\
\hline
  & \multicolumn{2}{c}{11.75} & \multicolumn{2}{c}{12.25} & \multicolumn{2}{c}{12.75} & \multicolumn{2}{c}{13.25} & \multicolumn{2}{c}{13.75} & \multicolumn{2}{c}{14.25} \\
 & SGP & Nexus & SGP & Nexus & SGP & Nexus & SGP & Nexus & SGP & Nexus & SGP & Nexus \\
\hline
8.5--9   &
$-9.62^{+0.07}_{-0.07}$ & -- &
$-9.71^{+0.23}_{-0.20}$ & -- &
$-9.73^{+0.06}_{-0.07}$ & $-9.60^{+0.06}_{-0.09}$ &
$-9.70^{+0.15}_{-0.13}$ & $-9.83^{+0.10}_{-0.09}$ &
$-9.63^{+0.05}_{-0.04}$ & $-9.61^{+0.06}_{-0.06}$ &
-- & -- \\
9--9.5   &
$-9.76^{+0.05}_{-0.06}$ & $-9.68^{+0.05}_{-0.05}$ &
$-9.75^{+0.03}_{-0.03}$ & $-9.86^{+0.04}_{-0.04}$ &
$-9.80^{+0.02}_{-0.02}$ & $-9.88^{+0.04}_{-0.05}$ &
$-9.82^{+0.02}_{-0.02}$ & $-9.83^{+0.03}_{-0.03}$ &
$-9.81^{+0.02}_{-0.02}$ & $-9.84^{+0.05}_{-0.05}$ &
$-9.84^{+0.05}_{-0.04}$ & -- \\
9.5--10  &
$-9.63^{+0.14}_{-0.16}$ & -- &
$-9.90^{+0.02}_{-0.02}$ & $-10.05^{+0.07}_{-0.06}$ &
$-9.88^{+0.01}_{-0.01}$ & $-9.92^{+0.05}_{-0.04}$ &
$-9.91^{+0.01}_{-0.01}$ & $-10.00^{+0.03}_{-0.03}$ &
$-9.99^{+0.02}_{-0.01}$ & $-10.04^{+0.05}_{-0.07}$ &
$-10.05^{+0.07}_{-0.06}$ & -- \\
10--10.5 &
-- & -- &
$-10.20^{+0.06}_{-0.06}$ & $-10.79^{+0.22}_{-0.24}$ &
$-10.19^{+0.02}_{-0.02}$ & $-10.49^{+0.16}_{-0.19}$ &
$-10.26^{+0.02}_{-0.02}$ & $-10.60^{+0.18}_{-0.15}$ &
$-10.42^{+0.03}_{-0.03}$ & $-10.93^{+0.13}_{-0.14}$ &
$-10.65^{+0.07}_{-0.09}$ & -- \\
10.5--11 &
-- & -- &
-- & -- &
$-10.63^{+0.04}_{-0.03}$ & $-11.06^{+0.19}_{-0.19}$ &
$-10.76^{+0.03}_{-0.03}$ & $-10.79^{+0.14}_{-0.13}$ &
$-10.90^{+0.04}_{-0.04}$ & $-11.13^{+0.14}_{-0.19}$ &
$-11.25^{+0.08}_{-0.07}$ & -- \\
11--11.5 &
-- & -- &
-- & -- &
-- & -- &
$-11.03^{+0.05}_{-0.08}$ & -- &
$-11.24^{+0.07}_{-0.07}$ & $-11.64^{+0.30}_{-0.29}$ &
$-11.18^{+0.23}_{-0.28}$ & -- \\
\hline
\end{tabular}
}
\end{table*}


\begin{table*}
\centering
\caption{Mean $\log \mathrm{sSFR}~[\log \mathrm{yr}^{-1}]$ of galaxies as a function of stellar mass for different halo mass bins for both the SGP and Nexus samples. Values are the bootstrapped medians, with uncertainties spanning the 16th and 84th percentiles. Bins with fewer than three galaxies are indicated by ``--''.}
\label{tab:stellar_halo_sSFR}
\resizebox{\textwidth}{!}{%
\begin{tabular}{c|cccccccccccc}
\hline
$M_{\mathrm{halo}}~[\log M_{\odot}]$ & \multicolumn{12}{c}{$M_{\mathrm{stellar}}$ [$\log M_{\odot}$]}\\
\hline
  & \multicolumn{2}{c}{8.75} & \multicolumn{2}{c}{9.25} & \multicolumn{2}{c}{9.75} & \multicolumn{2}{c}{10.25} & \multicolumn{2}{c}{10.75} & \multicolumn{2}{c}{11.25} \\
 & SGP & Nexus & SGP & Nexus & SGP & Nexus & SGP & Nexus & SGP & Nexus & SGP & Nexus \\
\hline
Field   &
$-9.72^{+0.03}_{-0.03}$ & $-9.71^{+0.05}_{-0.05}$ &
$-9.78^{+0.01}_{-0.01}$ & $-9.87^{+0.03}_{-0.03}$ &
$-9.90^{+0.00}_{-0.00}$ & $-10.02^{+0.03}_{-0.03}$ &
$-10.17^{+0.01}_{-0.01}$ & $-10.37^{+0.08}_{-0.09}$ &
$-10.57^{+0.02}_{-0.02}$ & $-10.85^{+0.15}_{-0.15}$ &
$-10.98^{+0.06}_{-0.06}$ & -- \\
11.5--12   &
$-9.62^{+0.07}_{-0.07}$ & -- &
$-9.76^{+0.05}_{-0.06}$ & $-9.68^{+0.05}_{-0.05}$ &
$-9.63^{+0.14}_{-0.16}$ & -- &
-- & -- &
-- & -- &
-- & -- \\
12--12.5   &
$-9.71^{+0.23}_{-0.20}$ & -- &
$-9.75^{+0.03}_{-0.03}$ & $-9.86^{+0.04}_{-0.04}$ &
$-9.90^{+0.02}_{-0.02}$ & $-10.05^{+0.07}_{-0.06}$ &
$-10.20^{+0.06}_{-0.06}$ & $-10.79^{+0.22}_{-0.24}$ &
-- & -- &
-- & -- \\
12.5--13   &
$-9.73^{+0.06}_{-0.07}$ & $-9.60^{+0.06}_{-0.09}$ &
$-9.80^{+0.02}_{-0.02}$ & $-9.88^{+0.04}_{-0.05}$ &
$-9.88^{+0.01}_{-0.01}$ & $-9.92^{+0.05}_{-0.04}$ &
$-10.19^{+0.02}_{-0.02}$ & $-10.49^{+0.16}_{-0.19}$ &
$-10.63^{+0.04}_{-0.03}$ & $-11.06^{+0.19}_{-0.19}$ &
-- & -- \\
13--13.5   &
$-9.70^{+0.15}_{-0.13}$ & $-9.83^{+0.10}_{-0.09}$ &
$-9.82^{+0.02}_{-0.02}$ & $-9.83^{+0.03}_{-0.03}$ &
$-9.91^{+0.01}_{-0.01}$ & $-10.00^{+0.03}_{-0.03}$ &
$-10.26^{+0.02}_{-0.02}$ & $-10.60^{+0.18}_{-0.15}$ &
$-10.76^{+0.03}_{-0.03}$ & $-10.79^{+0.14}_{-0.13}$ &
$-11.03^{+0.05}_{-0.08}$ & -- \\
13.5--14   &
$-9.63^{+0.05}_{-0.04}$ & $-9.61^{+0.06}_{-0.06}$ &
$-9.81^{+0.02}_{-0.02}$ & $-9.84^{+0.05}_{-0.05}$ &
$-9.99^{+0.02}_{-0.01}$ & $-10.04^{+0.05}_{-0.07}$ &
$-10.42^{+0.03}_{-0.03}$ & $-10.93^{+0.13}_{-0.14}$ &
$-10.90^{+0.04}_{-0.04}$ & $-11.13^{+0.14}_{-0.19}$ &
$-11.24^{+0.07}_{-0.07}$ & $-11.64^{+0.30}_{-0.29}$ \\
14--14.5   &
-- & -- &
$-9.84^{+0.05}_{-0.04}$ & -- &
$-10.05^{+0.07}_{-0.06}$ & -- &
$-10.65^{+0.07}_{-0.09}$ & -- &
$-11.25^{+0.08}_{-0.07}$ & -- &
$-11.18^{+0.23}_{-0.28}$ & -- \\
\hline
\end{tabular}
}
\end{table*}

\end{document}